# Alfvén Wave Dissipation in the Solar Chromosphere

Samuel D. T. Grant,[1, *] David B. Jess,[1, 2] Teimuraz V. Zaqarashvili,[3, 4, 5] Christian Beck,[6] Hector Socas-Navarro,[7, 8] Markus J. Aschwanden,[9] Peter H. Keys,[1] Damian J. Christian,[2] Scott J. Houston,[1] and Rebecca L. Hewitt[1]

[1] Astrophysics Research Centre, School of Mathematics and Physics, Queen's University Belfast, Belfast BT7 1NN, UK
[2] Department of Physics and Astronomy, California State University Northridge, Northridge, CA 91330, U.S.A.
[3] Space Research Institute, Austrian Academy of Sciences, Schmiedlstrasse 6, A-8042, Graz, Austria
[4] Abastumani Astrophysical Observatory at Ilia State University, 0162 Tbilisi, Georgia
[5] Institute of Physics, IGAM, University of Graz, Universitätsplatz 5, A-8010, Graz, Austria
[6] National Solar Observatory (NSO), Boulder, CO 80383, USA
[7] Instituto de Astrofísica de Canarias, Avda vía Láctea S/N, 38200, La Laguna, Tenerife, Spain
[8] Departamento de Astrofísica, Universidad de La Laguna, 38205, La Laguna, Tenerife, Spain
[9] Lockheed Martin, Solar and Astrophysics Laboratory, Org. A021S, Bldg. 252, 3251 Hanover St., Palo Alto, CA 94304, USA

*Correspondence addressed to samuel.grant@qub.ac.uk

**Magneto-hydrodynamic (MHD) Alfvén waves[1] have been a focus of laboratory plasma physics[2] and astrophysics[3] for over half a century. Their unique nature makes them ideal energy transporters, and while the solar atmosphere provides preferential conditions for their existence[4], direct detection has proved difficult as a result of their evolving and dynamic observational signatures. The viability of Alfvén waves as a heating mechanism relies upon the efficient dissipation and thermalization of the wave energy, with direct evidence remaining elusive until now. Here we provide the first observational evidence of Alfvén waves heating chromospheric plasma in a sunspot umbra through the formation of shock fronts. The magnetic field configuration of the shock environment, alongside the tangential velocity signatures, distinguish them from conventional umbral flashes[5]. Observed local temperature enhancements of 5% are consistent with the dissipation of mode-converted Alfvén waves driven by upwardly propagating magneto-acoustic oscillations, providing an unprecedented insight into the behaviour of Alfvén waves in the solar atmosphere and beyond.**

The solar surface hosts a web of diverse magnetic fields, from sunspots exhibiting sizes that dwarf the Earth, to dynamic bright grains only a few hundred km across. The magnetic nature of the Sun's atmosphere supports the plethora of MHD wave activity observed in recent years[6]. Such wave motion is predominantly generated near the surface of the Sun, with the creation of upwardly propagating MHD waves providing a conduit for the transportation of heat, from the vast energy reservoir of the solar photosphere, to the outermost extremities of the multi-million degree corona.

In comparison to other MHD modes, Alfvén waves are the preferred candidates for energy transport since they do not reflect or dissipate energy freely[3]. Observational studies have been limited by the challenging requirements on instrumentation needed to identify the Doppler line-of-sight (LOS) velocity perturbations and non-thermal broadening associated with Alfvén waves, thus there is only tentative evidence of their existence within the Sun's magnetized plasma[7-9]. Given the difficulties associated with resolving the intrinsic wave signatures, to date there has been no observational evidence brought forward to verify the dissipative processes associated with Alfvén waves. Theoretical studies have proposed multiple dissipation methods that would allow the embedded mechanical energy of Alfvén waves to be converted into localized heat[10,11]. Unfortunately, most act on unobservable scales, providing no clear signatures that can be identified with even the largest current solar telescopes. However, one distinct mechanism revolves around the formation of macroscopic shock fronts, which naturally manifest as a result of the propagation of waves through the solar atmosphere[12]. Shock behavior induced by slow magneto-acoustic waves is ubiquitously observed in sunspots, manifesting as umbral flashes[5] (UFs), giving rise to notable periodic intensity and temperature excursions[13]. Importantly, from a theoretical viewpoint, the shock dissipation of propagating wave energy also applies to Alfvén waves[14]. Here, large-amplitude linearly-polarized Alfvén waves induce their own shock fronts[15], or the associated ponderomotive force of propagating Alfvén waves leads to the excitation of resonantly shocked acoustic perturbations[16]. In this Letter, we present unique high-resolution observations and magnetic field extrapolations, combined with thermal inversions and MHD wave theory, to provide first-time evidence of Alfvén wave dissipation in the form of shock fronts manifesting close to the umbral boundary of a sunspot.

Intensity scans of the chromospheric Ca II 8542 Å spectral line at high spatial (71 km per pixel) and temporal (5.8 s) resolution were conducted for a 70 x 70 Mm$^2$ region centred around a large sunspot on 24 August 2014 using the Interferometric Bidimensional Spectrometer[17] (IBIS) at the Dunn Solar Telescope. Three-dimensional values for the vector magnetic field are derived through non-linear force-free extrapolations[18] applied to simultaneous magnetograms obtained by the Helioseismic and Magnetic Imager[19] (HMI) aboard the Solar Dynamics Observatory spacecraft. The resulting images (Fig. 1) highlight the connectivity between the magnetic fields and the structuring of the sunspot atmosphere. Intensity thresholding of running mean subtracted images of the chromospheric umbra allowed for the identification of 554,792 individual shock signals across the 135 minute duration of the dataset.

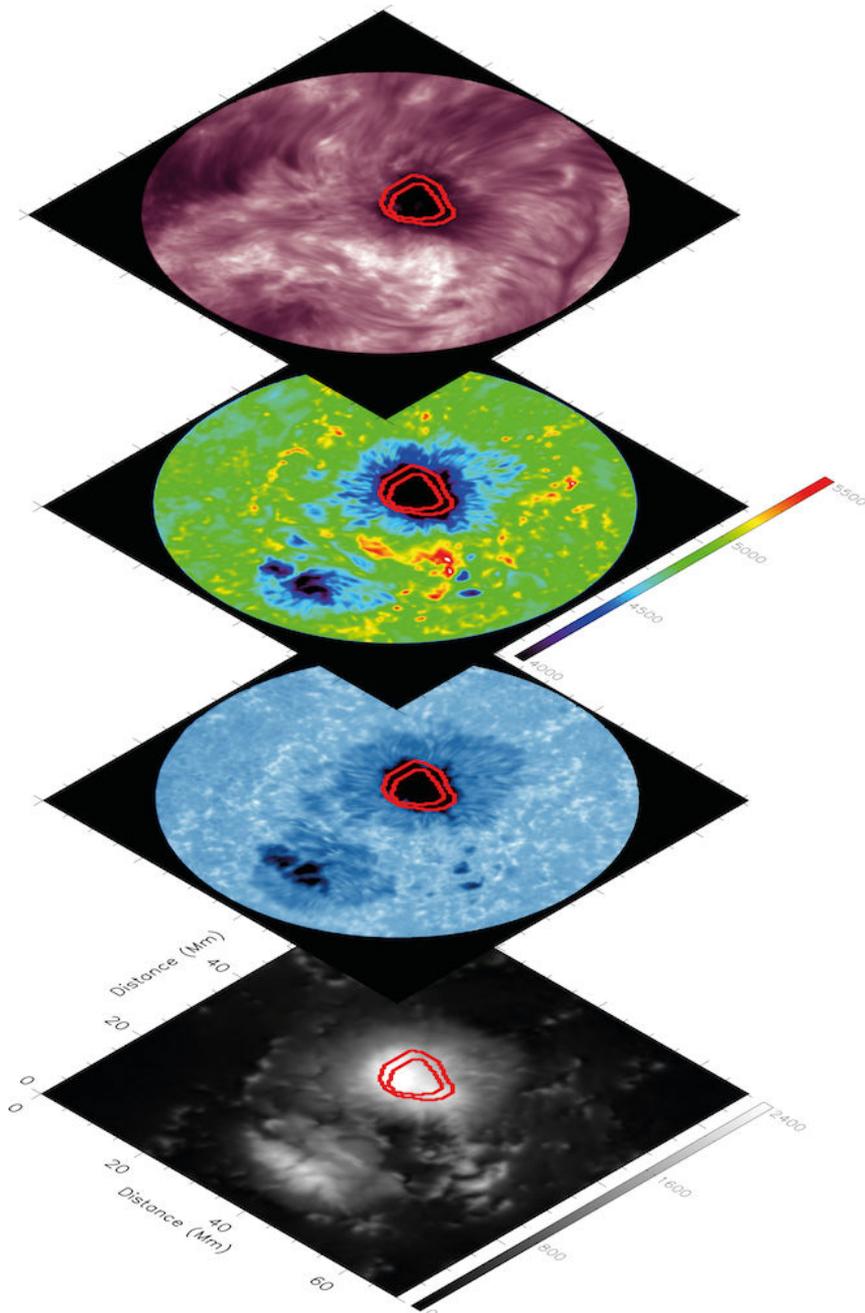

*Figure 1:* **The building blocks of the magnetized solar atmosphere observed on 2014 August 24.** *Co-spatial images revealing the structure of the sunspot at 13:00 UT on 24 August 2014. The lower panel shows the magnitude of the photospheric magnetic field from HMI, revealing high umbral field strengths (colorbar relates to the field strengths in Gauss). The subsequent panel is an image taken from the blue wing of the Ca II 8542 Å spectral line, displaying the photospheric representation of the sunspot. Above this is the photospheric plasma temperature of the region derived from CAISAR at log ($\tau_{500nm}$) ~ -2 (or ~250 km above the photosphere), showing the clear temperature distinction between the umbra, penumbra and surrounding quiet Sun (colorbar in units of Kelvin). The upper panel shows the chromospheric core of the Ca II 8542 Å spectral line, highlighting the strong intensity gradient between the umbra and penumbra at these heights. In each of these images, the red contours represent the inner and outer boundaries of the plasma-$\beta$ = 1 region at the height where shocks first begin to manifest (~250 km), where magneto-acoustic and Alfvén waves can readily convert.*

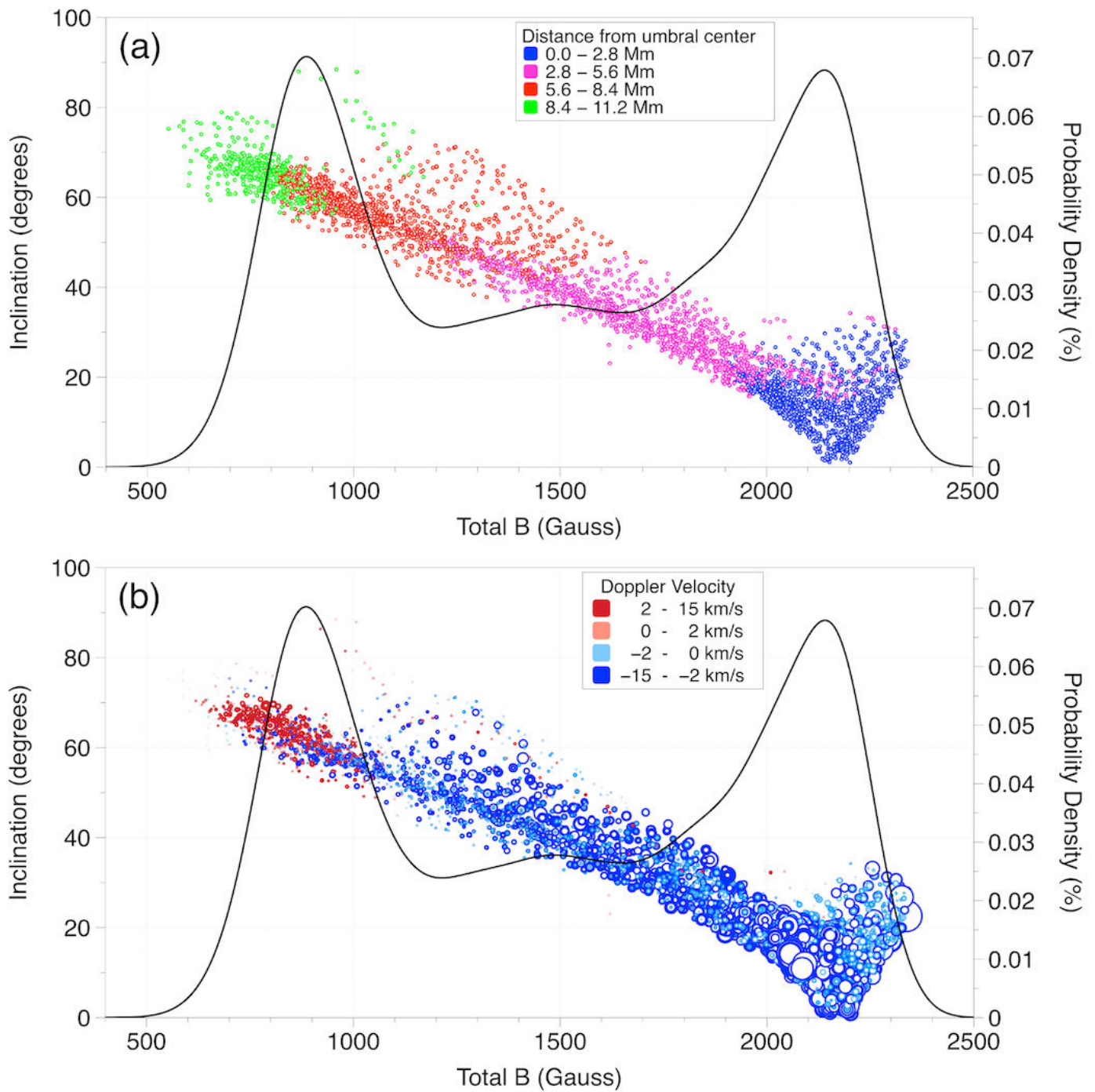

*Figure 2:* **A statistical insight into the magnetic, velocity and occurrence relationships between shock phenomena in a sunspot umbra.** *The inclination of the magnetic field as a function of the total magnetic field strength for shock pixels at their point of formation, where an inclination of 0° represents an upwardly orientated vertical field. The color scheme of the data points denotes their distance from the centre of the umbral core in Mm (panel **a** and the LOS velocities of the resultant shock emission profiles, where positive values represent red-shifted plasma motion away from the observer, while negative values indicate blue-shifted plasma towards the observer (panel **b**). The diameters of the data points in the lower panel represent a visualization of the relative temperature increases of shocked plasma above their local quiescent background temperatures, where the largest circles represent significant temperature enhancements (up to a maximum of ~20%) that are synonymous with higher magnetic field strengths when compared to the smaller temperature enhancements associated with lower magnetic field strengths. The data points are over-plotted with the probability density function of shock occurrence as a function of the total magnetic field strength.*

The sunspot displays a classical structure, providing two preferential locations for plasma shock formation in the umbra: (1) in the vicinity of strong, vertical magnetic fields near the umbral centre-of-mass, and (2) in the presence of weaker, inclined fields towards the outer boundary of the umbra (Fig. 2). The first population is attributed to UFs, whereby the near-vertical propagation of magneto-acoustic waves across multiple density scale heights promotes their efficient steepening into shocks[20]. The second population, whose intensity excursions are formed along inclined magnetic fields that channel waves almost horizontally (inclinations of approximately 70 - 80 degrees), provides the first indication of Alfvén wave shock formation. Alfvén shocks are predicted to form in regions with high negative Alfvén speed gradients[21], which is fulfilled by the volume expansion of the magnetic field lines and the highly inclined environment in which they exist, hence minimizing the effects of density stratification and promoting a negative gradient in the associated Alfvén speed.

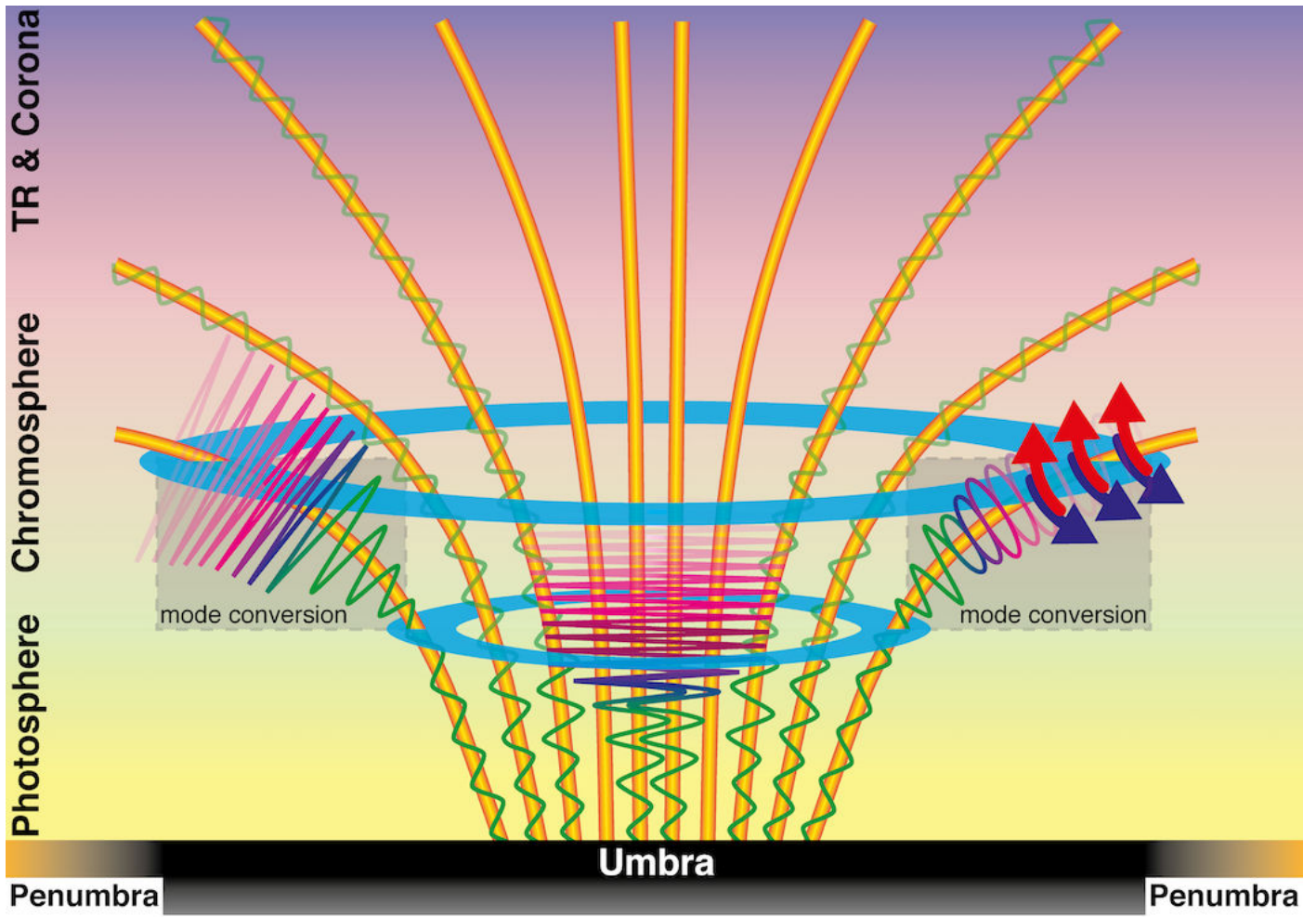

*Figure 3: **A cartoon representation of a sunspot umbral atmosphere demonstrating a variety of shock phenomena.** A side-on perspective of a typical sunspot atmosphere, showing magnetic field lines (orange cylinders) anchored into the photospheric umbra (bottom of image) and expanding laterally as a function of atmospheric height. The light blue annuli highlight the lower and upper extents of the mode conversion region for the atmospheric heights of interest. The mode conversion region on the left-hand side portrays a schematic of non-linear Alfvén waves resonantly amplifying magneto-acoustic waves, increasing the shock formation efficiency in this location. The mode conversion region on the righthand side demonstrates the coupling of upwardly propagating magneto-acoustic oscillations (the sinusoidal motions) into Alfvén waves (the elliptical structures), which subsequently develop tangential blue- and red-shifted plasma during the creation of Alfvén shocks. The central portion represents the traditional creation of UFs that result from the steepening of magneto-acoustic waves as they traverse multiple density scale heights in the lower solar atmosphere. Image not to scale.*

The extrapolated vector magnetic fields are employed with the complementary CAlcium Inversion using a Spectral Archive[22] (CAISAR) code, producing maps of magnetic and thermal pressure that are used to establish the physical locations where the ratio of these parameters, the plasma-$\beta$, equals unity (Fig. 1). Here, the plasma-$\beta$ = 1 region is analogous to the locations where the characteristic velocities of slow and fast waves are equal. Thus, ubiquitous magneto-acoustic waves throughout the umbra are capable of converting into Alfvén modes through the process of mode conversion[11,23]. This provides a bulk generation of Alfvén waves that can form both Alfvén and resonantly driven shocks under the correct atmospheric conditions. The magnetic topologies representative of the outermost boundary of the sunspot umbra provide such an environment where a steep negative Alfvén speed gradient is encountered by the propagating waves. UF shock formation in this regime is drastically suppressed due to the heavily inclined magnetic field lines providing substantially reduced density stratification along the paths of wave propagation.

Examination of the Doppler LOS velocities of the shocked plasma further distinguish between UFs near the sunspot core and the Alfvén shocks at the periphery of the umbra. The LOS plasma velocities associated with the UF population are blue-shifted and display characteristic 'sawtooth' spectral profiles throughout their temporal evolution, consistent with the established morphology of UFs[24]. The second population, which dominates the outer umbral perimeter where the plasma-β equals unity, displays an intermingling of red- and blue-shifted plasma moving perpendicularly to the wave vector (Fig. 2) during the onset of plasma shock events. This is in stark contrast to conventional UFs, and provides further direct evidence of Alfvén shocks, since large velocity excursions perpendicular to the vector magnetic field are representative of Alfvén waves undergoing non-linear processes during the creation of shocks[21] (Fig. 3). Furthermore, the observed red-shifts indicate that approximately 70% of the observed shocks are due to the direct steepening of Alfvén waves, which is depicted in the righthand side of Fig. 3.

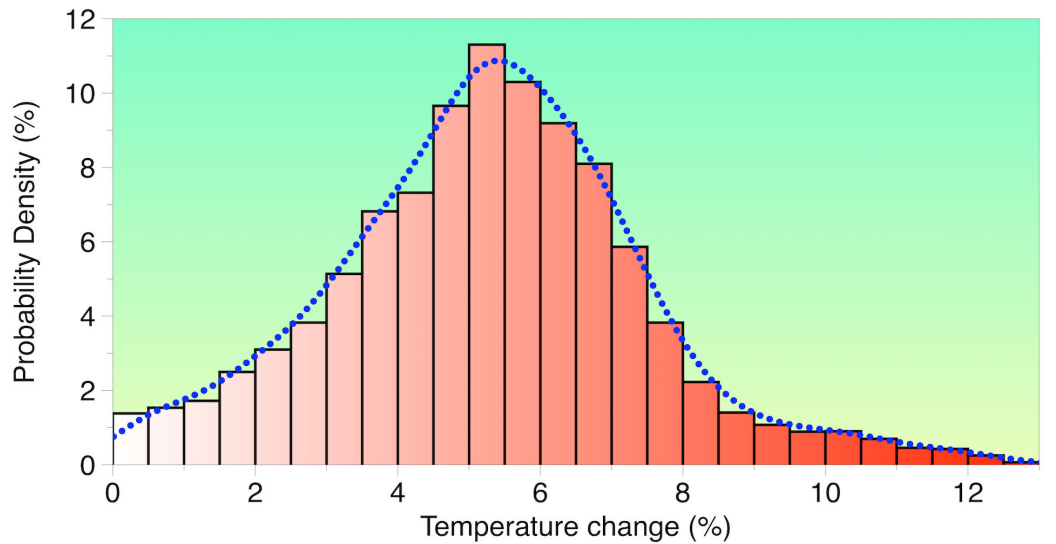

Temporal analysis of the CAISAR inversions unveils how the strong dissipative thermal effects of the Alfvén shocks first appear just above the temperature minimum at $log\ (\tau_{500nm})$ – 2 (~250 km above the solar surface), where $\tau_{500nm}$ represents the optical depth of the plasma at a wavelength of 500 nm. The greatest thermal perturbations occur between optical depths of -5.3 < $log\ (\tau_{500nm})$ < -4.6 (~750 – 1100 km), corresponding to the low chromosphere[25].

*Figure 4: **The distribution of temperature enhancements resulting from Alfvén shocks close to the umbral boundary.** The probability density of percentage temperature perturbations derived from Alfvén shocks with respect to their average background temperature. The blue dotted line is a smoothed fit to the histogram, showing a clear peak at approximately 5%, which equates to the conversion of mechanical Alfvén wave flux into thermalized shock energy.*

Through comparison with the average background temperature, the Alfvén shocks are found to heat the localized plasma by ~5% (Fig. 4). By comparison, the ubiquitous slow magneto-acoustic shocks (UFs) identified at the cooler core of the sunspot umbra exhibit temperature increases on the order of 20% (lower panel of Fig. 2), consistent with previous estimations[13].

The wealth of high-resolution data allows for the calculation of the Alfvén wave energy flux that subsequently steepens into shocks. The energy flux, $E_A$, of an Alfvén wave can be expressed as,

$$E_A = \rho\, v^2\, c_A$$

where $\rho = 3.57 \pm 1.30 \times 10^{-7}$ kg m$^{-3}$ is the density found from non-local thermodynamic equilibrium inversions[26], and $c_A = 6.2 \pm 0.5$ km s$^{-1}$ is the local Alfvén speed computed from the vector magnetic field extrapolations, with both values computed at the atmospheric height and outer umbral location synonymous with the Alfvén shock detections ($log\ (\tau_{500nm})$ ~ -5 within the plasma-β = 1 isocontours in Fig. 1). The calculated Alfvén speed is also consistent with modern statistical seismological findings[27]. A velocity amplitude, $v = 2.2 \pm 0.6$ km s$^{-1}$, is computed from the average absolute velocity values related to the Alfvén wave population displayed in Fig. 2. These measurements, derived directly from the observations, output an Alfvén energy flux of $E_A = 10.7 \pm 5.7$ kW m$^{-2}$. This energy is substantially less than the 20 kW m$^{-2}$ observed for upwardly propagating magneto-acoustic waves generated in photospheric sunspot umbrae[28]. Such a differential is not unexpected, as only a portion of the photospheric flux will mode convert into Alfvén waves within the $\beta = 1$ layer[29]. Furthermore, partially ionized atoms that make up the cool umbral chromosphere will naturally absorb a significant portion of the available energy through ionization and ion-neutral processes above the temperature minimum[30]. Hence, only a fraction of the initial energy will directly contribute to the observed temperature increase of ~5%, thus remaining consistent with the scenario of magneto-acoustic mode conversion leading to the steepening of Alfvén waves into shock fronts.

Here, we show the dissipation of chromospheric Alfvén waves into thermal energy for the first time. We reveal how high-resolution observations provide the direct detection of these signatures, in the form of identifying the necessary plasma=$\beta = 1$ equipartition region and revealing the co-spatial negative Alfvén speed gradients and associated tangential Doppler velocities synonymous with Alfvén shock dissipation, to provide first light on an enigmatic problem within astrophysics that has remained elusive for over half a century. With the viability of Alfvén waves to heat localized plasma confirmed, future studies can build upon our novel findings and examine the consequences of Alfvén shocks in greater detail using spectro-polarimetric data and advanced inversion techniques. Looking further ahead, the capabilities of the upcoming 4m Daniel K. Inouye Solar Telescope, set for first light in 2019, will unveil the fine structure of these shocks on an unprecedented scale, allowing the role Alfvén shocks play in global solar atmospheric heating to be fully assessed.

**Data Availability**

The data used in this paper are from the observing campaign entitled *'Nanoflare Activity in the Lower Solar Atmosphere'* (NSO-SP proposal T1020; Principle Investigator: D. B. Jess), which employed the ground-based Dunn Solar Telescope, USA, during August 2014. Additional supporting observations were obtained from the publicly available NASA's Solar Dynamics Observatory (https://sdo.gsfc.nasa.gov) data archive, which can be accessed via http://jsoc.stanford.edu/ajax/lookdata.html. The ground-based data obtained during 2014 August 24 is several TB in size, and cannot be hosted on a public server. However, all data supporting the findings of this study are available directly from the authors on request.

**Acknowledgments**

S.D.T.G. and S.J.H. thank the Northern Ireland Department for Employment and Learning (now the Northern Ireland Department for the Economy) for the awards of Ph.D. studentships. D.B.J. wishes to thank the UK Science and Technology Facilities Council for the award of an Ernest Rutherford Fellowship alongside a dedicated Research Grant. S.D.T.G and D.B.J. also wish to thank Invest NI and Randox Laboratories Ltd. for the award of a Research & Development Grant (059RDEN-1) that allowed this work to be undertaken. P.H.K. is grateful to the Leverhulme Trust for the award of an Early Career Fellowship. The NSO is operated by the Association of Universities for Research in Astronomy under cooperative agreement with the National Science Foundation. The magnetic field measurements employed in this work are courtesy of NASA/SDO and the AIA, EVE, and HMI science teams.

**References**

1. Alfvén, H. Magneto hydrodynamic waves, and the heating of the solar corona. *MNRAS* **107**, 211 (1947).
2. Gekelman, W., Vincena, S., Leneman, D. & Maggs, J. Laboratory experiments on shear Alfvén waves and their relationship to space plasmas. *J. Geophys. Res.* **102**, 7225-7236 (1997).
3. Mathioudakis, M., Jess, D. B. & Erdélyi, R. Alfvén waves in the solar atmosphere. From theory to observations. *Space Sci. Rev.* **175**, 1-27 (2013).

4. Morton, R. J., Verth, G., Fedun, V., Shelyag, S. & Erdélyi, R. Evidence for the photospheric excitation of incompressible chromospheric waves. *Astrophys. J.* **768**, 17 (2013).
5. Beckers, J. M. & Tallant, P. E. Chromospheric inhomogeneities in sunspot umbrae. *Sol. Phys.* **7**, 351-365 (1969).
6. Jess, D. B. et al. Multiwavelength studies of MHD waves in the solar chromosphere. An overview of recent results. *Space Sci. Rev.* **190**, 103 (2015).
7. Tomczyk, S. et al. Alfvén waves in the solar corona. *Science* **317**, 1192 (2007).
8. De Pontieu, B. et al. Chromospheric Alfvén waves strong enough to power the solar wind. *Science* **318**, 1574 (2007).
9. Jess, D. B. et al. Alfvén waves in the lower solar atmosphere. *Science* **323**, 1582 (2009).
10. Heyvaerts, J. & Priest, E. R. Coronal heating by phase-mixed shear Alfvén waves. *Astron. Astrophys.* **117**, 220-234 (1983).
11. Bogdan, T. J. et al. Waves in the magnetized solar atmosphere. II. Waves from localized sources in magnetic flux concentrations. *Astrophys. J.* **599**, 626-660 (2003).
12. Schwarzschild, M. On noise arising from the solar granulation. *Astrophys. J.* **107**, 1 (1948).
13. de la Cruz Rodríguez, J., Rouppe van der Voort, L., Socas-Navarro, H. & van Noort, M. Physical properties of a sunspot chromosphere with umbral flashes. *Astron. Astrophys.* **556**, A115 (2013).
14. Arber, T. D., Brady, C. S. & Shelyag, S. Alfvén wave heating of the solar chromosphere: 1.5D models. *Astrophys. J.* **817**, 94 (2016).
15. Montgomery, D. Development of hydromagnetic shocks from large-amplitude Alfvén waves. *Phys. Rev. Lett.* **2**, 36-37 (1959).
16. Hada, T. Evolution of large amplitude Alfvén waves in the solar wing with beta approximately 1. *Geophys. Rev. Lett.* **20**, 2415- 2418 (1993).
17. Cavallini, F. IBIS: A new post-focus instrument for solar imaging spectroscopy. *Sol. Phys.* **236**, 415-439 (2006).
18. Wiegelmann, T. Nonlinear force-free modelling of the solar coronal magnetic field. *Journal of Geophysical Research (Space Physics)* **113**, A03S02 (2008).
19. Schou, J. et al. Design and ground calibration of the helioseismic and magnetic imager (HMI) instrument on the solar dynamics observatory (SDO). *Sol. Phys.* **275**, 229 (2012).
20. Bard, S. & Carlsson, M. Radiative hydrodynamic simulations of acoustic waves in sunspots. *Astrophys. J.* **722**, 888 (2010).
21. Hollweg, J. V., Jackson, S. & Galloway, D. Alfvén waves in the solar atmosphere. III - Nonlinear waves on open flux tubes. *Sol. Phys.* **75**, 35-61 (1982).
22. Beck, C., Choudhary, D. P., Rezaei, R. & Louis, R. E. Fast inversion of solar Ca II spectra. *Astrophys. J.* **798**, 100 (2015).
23. Khomenko, E. & Cally, P. S. Numerical simulations of conversion to Alfvén waves in sunspots. *Astrophys. J.* **746**, 68 (2012).
24. Rouppe van der Voort, L. H. M., Rutten, R. J., Sütterlin, P., Sloover, P. J. & Krijger, J. M. La Palma observations of umbral flashes. *Astron. Astrophys.* **403**, 277-285 (2003).
25. Maltby, P. et al. A new sunspot umbral model and its variation with the solar cycle. *Astrophys. J.* **306**, 284-303 (1986).
26. Socas-Navarro, H., de la Cruz Rodríguez, J., Asensio Ramos, A., Trujillo Bueno, J. & Ruiz Cobo, B. An open-source, massively parallel code for non-LTE synthesis and inversion of spectral lines and Zeeman-induced Stokes profiles. *Astron. Astrophys.* **577**, A7 (2015).
27. Cho, I.-H. et al. Determination of the Alfvén speed and plasma-beta using the seismology of sunspot umbra. *Astrophys. J.* **837**, L11 (2017).
28. Kanoh, R., Shimizu, T. & Imada, S. Hinode and IRIS observations of the magnetohydrodynamic waves propagating from the photosphere to the chromosphere in a sunspot. *Astrophys. J.* **831**, 24 (2016).
29. Cally, P. S. & Moradi, H. Seismology of the wounded Sun. *MNRAS* **435**, 2589-2597 (2013).

30. Fontenla, J. M., Curdt, W., Haberreiter, M., Harder, J. & Tian, H. Semiempirical models of the solar atmosphere. III. Set of non- LTE models for the far-ultraviolet/extreme-ultraviolet irradiance computation. *Astrophys. J.* **707**, 482-502 (2009)

# Alfvén Wave Dissipation in the Solar Chromosphere
# Supplementary Material

## 1. Observations

The data presented here is an observational sequence obtained during 13:00 - 15:15 UT on 2014 August 24 with the Dunn Solar Telescope at Sacramento Peak, New Mexico. The telescope was pointed to active region NOAA 12146, positioned at heliocentric co-ordinates (496", 66"), or N10W32 in the conventional heliographic co-ordinate system, with exceptional seeing conditions throughout. The Interferometric BIdimensional Spectrometer[S1] (IBIS) was utilized to sample the Ca II absorption profile at 8542.12 Å with 27 non-equidistant wavelength steps employed (Fig. S1). The IBIS instrument imaged a 97" x 97" (approximately 70 x 70 Mm$^2$) region of the solar disk centred around a prominent sunspot. A spatial sampling of 0."098 (71 km) per pixel and temporal cadence of 5.8 s per full scan were utilized to produce a total of 1336 spectral imaging scans. A whitelight camera, synchronized with the narrowband channel, was also utilized to enable post-processing of the narrowband images, alongside the implementation of high-order adaptive optics[S2] to improve quality.

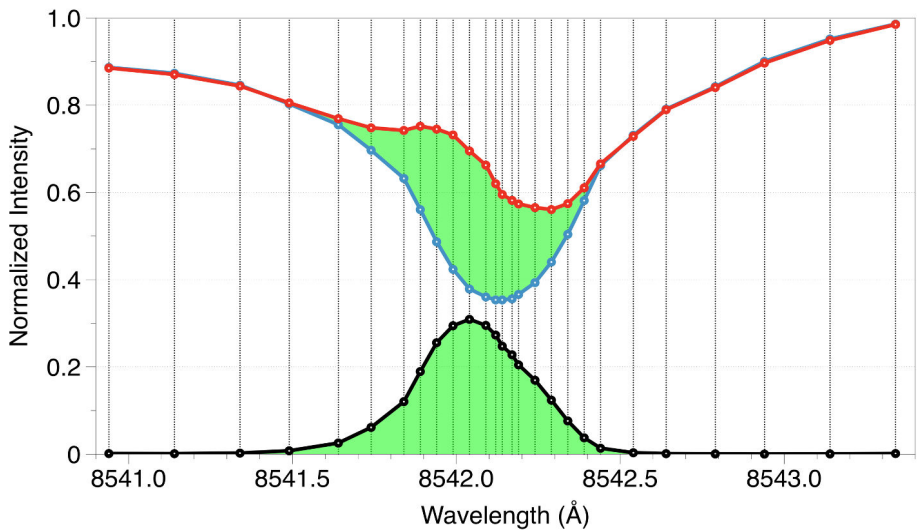

Figure 1: **The spectral profiles associated with quiescent and shocking umbral plasma**. The average quiescent Ca II 8542 Å umbral spectral profile (blue line) captured by IBIS, over-plotted with the average spectral profile corresponding to pixels demonstrating magneto-acoustic shock behavior (red line), where enhanced blue-shifted emission is clearly evident. The resultant profile, created by subtracting the average quiescent Ca II 8542 Å umbral spectral profile from the average profile corresponding to shock formation is displayed as a black line. What remains is an emission profile that allows for a clearer estimation of the shock Doppler velocity. The vertical dotted lines represent the wavelength sampling of IBIS used to cover the Ca II 8542 Å spectral line out to ± 1.2Å from line core.

Data from the Helioseismic and Magnetic Imager[S3] (HMI), aboard the Solar Dynamics Observatory[S4] (SDO), was used to provide simultaneous vector magnetograms of the region of interest. The simultaneous Milne-Eddington vector magnetograms were retrieved with a time cadence of 720 s and a two-pixel spatial resolution of 1."0 (725 km). Alongside this, one contextual HMI 6173 Å continuum image was acquired, corresponding to the beginning of the observing run, which was used for subsequent co-alignment of the IBIS data set. To this end, a sub-field was extracted from a near-simultaneous HMI continuum image and co-aligned to the IBIS whitelight channel. Cross-correlation techniques were used to provide excellent co-alignment accuracy, making pixel-by-pixel comparisons between the thermal inversions and the non-linear force-free magnetic field extrapolations possible.

## 2. Magnetic Field Extrapolations

Magnetic field extrapolations were conducted on the outputs of the Very Fast Inversion of the Stokes Vector[S5] (VFISV) algorithm applied to the HMI observations. The resultant heliocentric data was converted into a heliographic projection[S6], providing magnetic field components parallel, $B_x$ and $B_y$, and perpendicular, $B_z$, to the solar surface. At this point, a non-linear force-free field extrapolation code[S7] was employed. The extrapolations were applied to the derived heliographic components[S8,S9], producing information related to the vector magnetic field as a function of atmospheric height (Fig. S2). The final outputs of the extrapolations were converted into key parameters for the present study, namely the transverse ($B_{trans}$) and total ($B_{tot}$) magnetic field strengths, in addition to the inclination and azimuth angles, all as a function of atmospheric height.

## 2A. Validation of the Force-Freeness of the Sunspot Atmosphere

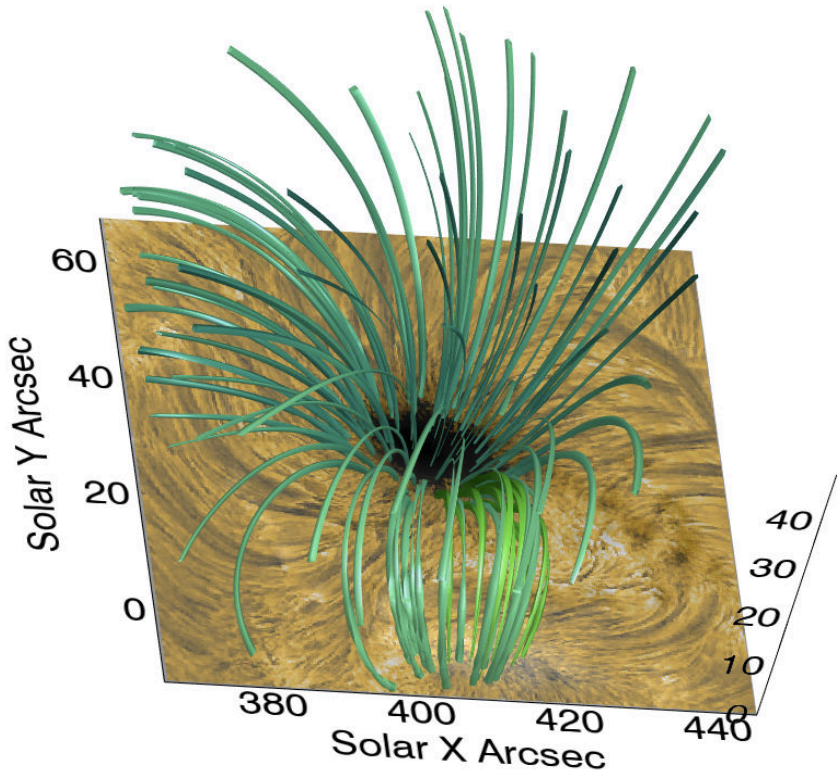

Figure 2: **The magnetic structuring of the 2014 August 24 sunspot atmosphere.** *A three-dimensional visualization of the extrapolated magnetic field lines (green contours) as they extend upwards through the solar atmosphere. Only magnetic field lines that originate within the umbra are displayed to highlight the magnetic topology intrinsic to the lower atmosphere of the sunspot under investigation. A chromospheric H$\alpha$ reference image, obtained with the Hydrogen-Alpha Rapid Dynamics camera[S15] (HARDcam) system, is displayed for contextual purposes since its field-of-view is larger than that provided by IBIS. The vertical axis is in extrapolated pixels, where 1 pixel corresponds to 362.5 km.*

Caveats are often applied to the outputs (particularly to the strengths and inclination angles) of non-linear force-free magnetic field extrapolations to account for uncertainties arising when the field lines are extended through the potentially volume-filling chromosphere and density discontinuous transition region[S10]. As the analysis presented here hinges upon the reliability of non-linear force-free extrapolations pertaining to the sunspot umbra, it naturally becomes a requirement to validate the degree of force-freeness embodied by the sunspot atmosphere.

Recent investigations[S11] have demonstrated that sunspot umbral atmospheres, even within the more challenging confines of chromospheric layers, are capable of being dominated by force-free conditions. However, this has not been universally observed across all sunspots studied[S12], therefore consideration of each chromospheric sunspot in isolation must be made to verify whether force-free conditions exist. Here, the plasma is defined as being 'force-free' when the local magnetic pressure dominates the plasma pressure, providing $\beta < 1$ (see Section 9). To better quantify the degree of force-freeness for active region NOAA 12146, a vertical current approximation magnetic field extrapolation code[S13] was used in conjunction with automated feature detection and isolation software[S14] applied to complementary H$\alpha$ observations obtained with the Hydrogen-Alpha Rapid Dynamics camera[S15] (HARDcam). A reference HARDcam H$\alpha$ image was utilized as this instrument provided the largest simultaneous chromospheric field-of-view size of approximately 180" x 180" (approximately 130 x 130 Mm$^2$), while still maintaining a diffraction-limited spatial sampling of 0."109 (80 km) per pixel. This allowed the magnetic field topology to be forward fitted to the identified solar structures, thus minimising misalignment angles and more accurately representing the true orientation of the magnetic field in the solar chromosphere. The established median misalignment angles in two dimensions ($\mu_2 = 5.9°$) and three dimensions ($\mu_3 = 10.7°$) are consistent with previous stereoscopic comparisons employing multiple datasets covering a variety of vantage points[S16] (Fig. S3). Thus, the resulting topology can be trusted to be an accurate representation of the magnetic field geometry within the solar chromosphere and beyond.

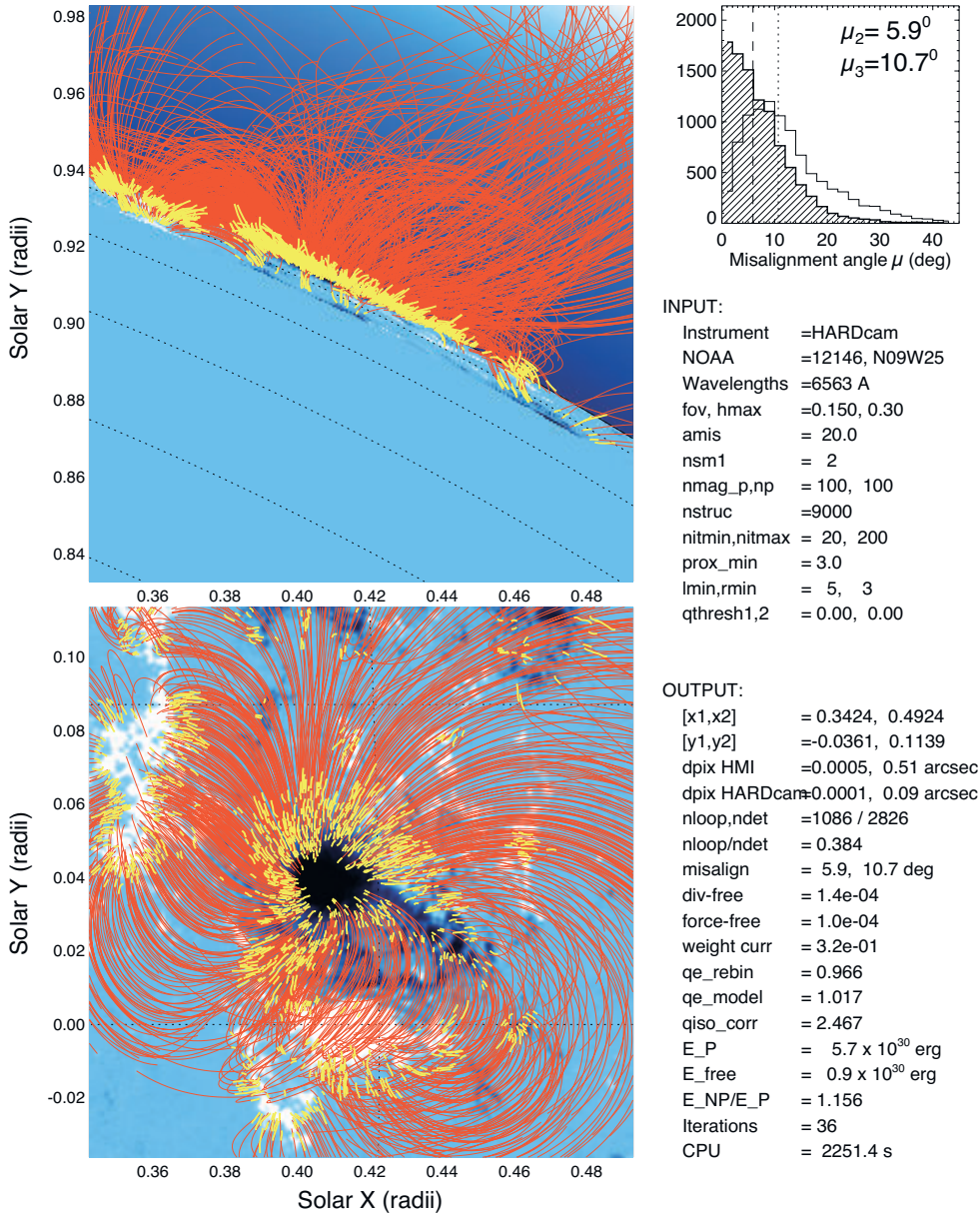

Figure 3: **The forward fitting of magnetic field lines from the 2014 August 24 active region.** *The automated curvi-linear feature traces from the chromospheric HARDcam Hα reference image obtained at 6563 Å are shown using yellow curves, overlaid on the best-fit solutions of the magnetic field model[S13] using red curves, and the observed SDO/HMI magnetogram (blue background image), from the line-of-sight view in the (x, y)-plane (bottom panel) and the orthogonal projection in the (x, z)-plane (top panel). A histogram of the 2D ($\mu_2$) and 3D ($\mu_3$) misalignment angles and various input and output parameters are shown at the right-hand side.*

With the three-dimensional magnetic field known to a high degree of precision, the degree of force-freeness can easily be modeled. Employing an established stratified model sunspot umbral atmosphere (model 'M')[S17], the computed plasma pressures can be compared directly to the magnetic pressures extracted from the forward-fitted magnetic field geometries. The ratio of these two parameters allows the calculation of the plasma-β (see Section 9) to be uncovered as a function of geometric height (Fig. S4). As one would expect, strong magnetic fields synonymous with umbral cores have exceptionally low plasma-β values, which remain ≤ 1 (i.e., predominantly force-free) across all atmospheric heights. Some weaker magnetic fields exist at chromospheric heights, which naturally result in a plasma pressure dominated regime (i.e., plasma-β > 1). However, for the sunspot under present investigation, the dark umbral core demonstrates mostly force-free (i.e., plasma-β < 1) conditions. This independent approach verifies the force-freeness of the sunspot umbral atmosphere, thus validating the use of the non-linear force-free extrapolation code[S7] employed for subsequent study.

## 2B. Examination of the Free Energy in Active Region NOAA 12146

Small-scale brightenings within active regions may arise from numerous physical processes occurring within the solar atmosphere. Of particular note are those related to umbral flashes[S18], umbral/penumbral jets[S19-S22], and reconnection phenomena in the form of microflares[S23,S24]. Therefore, it is important to ensure that the brightenings detected in the sunspot umbra are not related to jets and/or magnetic reconnection. The spectral signatures associated with the detected umbral events clearly segregate them from jet and reconnection phenomena (Fig. S5). Very clear blue-shifted brightenings are observed, with accelerations on the order of ≈4.40 km s$^{-2}$. These are followed by cooling red-shifted plasma falling back towards the solar surface, obeying the characteristic gravitational acceleration of the Sun (≈0.25 km s$^{-2}$). Such classic `sawtooth' signatures are consistent with previously observed magneto-acoustic shock phenomena[S25], confirming the umbral brightenings investigated here represent shocked plasma.

However, in order to more quantifiably exclude reconnection phenomena as a cause of the detected brightenings, we investigate the free magnetic energy contained within the entire active region. The free magnetic energy is defined as the excess energy embodied by the magnetic fields, above that which would be present under the most simplified magnetic configuration (i.e., a completely potential field). Now that the force-freeness of the sunspot atmosphere has been verified (see Section 2A), we apply a code[S26] to the extrapolated magnetic fields that is designed to calculate the magnetic energy and relative magnetic helicity budgets in solar active regions. At the beginning of the time series, a free magnetic energy of 7.9 ± 0.4 x 10$^{29}$ erg is calculated, which rises to 8.1 ± 0.5 x 10$^{29}$ erg at the end of the 135 minute observing sequence. This is also consistent with an independent measure of the free magnetic energy using vertical current approximation non-linear force-free magnetic field extrapolations[S13] (Fig. S3). A rise in free energy suggests that the magnetic field is becoming increasingly non-potential, perhaps due to developing helicity in the corresponding magnetic loop structures. However, with errors considered[S27], the rise in free energy is relatively insignificant when compared to the large changes in free magnetic energy leading up to flaring events[S28]. Importantly, there is no appreciable decline in the total amount of free energy, which removes widespread reconnection phenomena as a potential cause of the consistently visible plasma brightenings in the sunspot umbra. Therefore, the clear 'sawtooth' spectral signatures, coupled with a consistently measured non-decreasing free magnetic energy, suggests that the quasi-periodic brightenings observed in the sunspot umbra are a clear manifestation of plasma shock phenomena.

## 2C. Preparing the Magnetic Field Extrapolations for Analysis

At this stage, the magnetic maps were interpolated from the HMI spatial sampling onto the IBIS plate scale. Given the reduced temporal resolution of the HMI data from which the field extrapolations were derived, each IBIS frame corresponding to a new shock event was associated with the magnetic field information closest in time. This is an accurate assumption, since an inspection of the extrapolated fields show minimal structural evolution during the time of the observations; most likely a result of capturing an inactive sunspot group. The atmospheric height values of the extrapolations that are of interest for the present study were constrained through consideration of the IBIS Ca II 8542 Å contribution function[S29], which identifies the formation height of the line core between 500 – 1500 km. Thus, the extrapolated magnetic field values produced within the height interval of approximately 360 -1450 km (i.e., four extrapolation elements along the z-axis), were averaged to form two-dimensional maps of $B_{tot}$, $B_{trans}$, inclination and azimuth angles best matched to the IBIS observations. The height range used for the averaging is also consistent with the optical depth-based geometric heights extracted from sunspot model atmospheres[S17]. Furthermore, the averaging over 1090 km in atmospheric height is valid due to the extrapolated fields show minimal geometric changes across this narrow interval.

## 2D. The Veracity of the Extrapolated Magnetic Field

The results presented in this study are complemented by the identification of the underlying umbral magnetic field geometry. Therefore, confidence in the applied techniques and resulting outputs is vital. Initially, it was necessary to confirm that the magnetic fields remain uniform throughout the time period of the observations, and that no field re-alignments due to, e.g., reconnection events were taking place, which would add ambiguity to any magnetic field inference. To address this, the free magnetic energy embodied by the active region was quantified across the duration of the time series, with a small increase in the free energy observed. An increase in the free magnetic energy confirms that no discernible macroscopic reconnection events are taking place within the dataset, highlighting that no dynamic magnetic field evolutions occur within the active region over the period of observation.

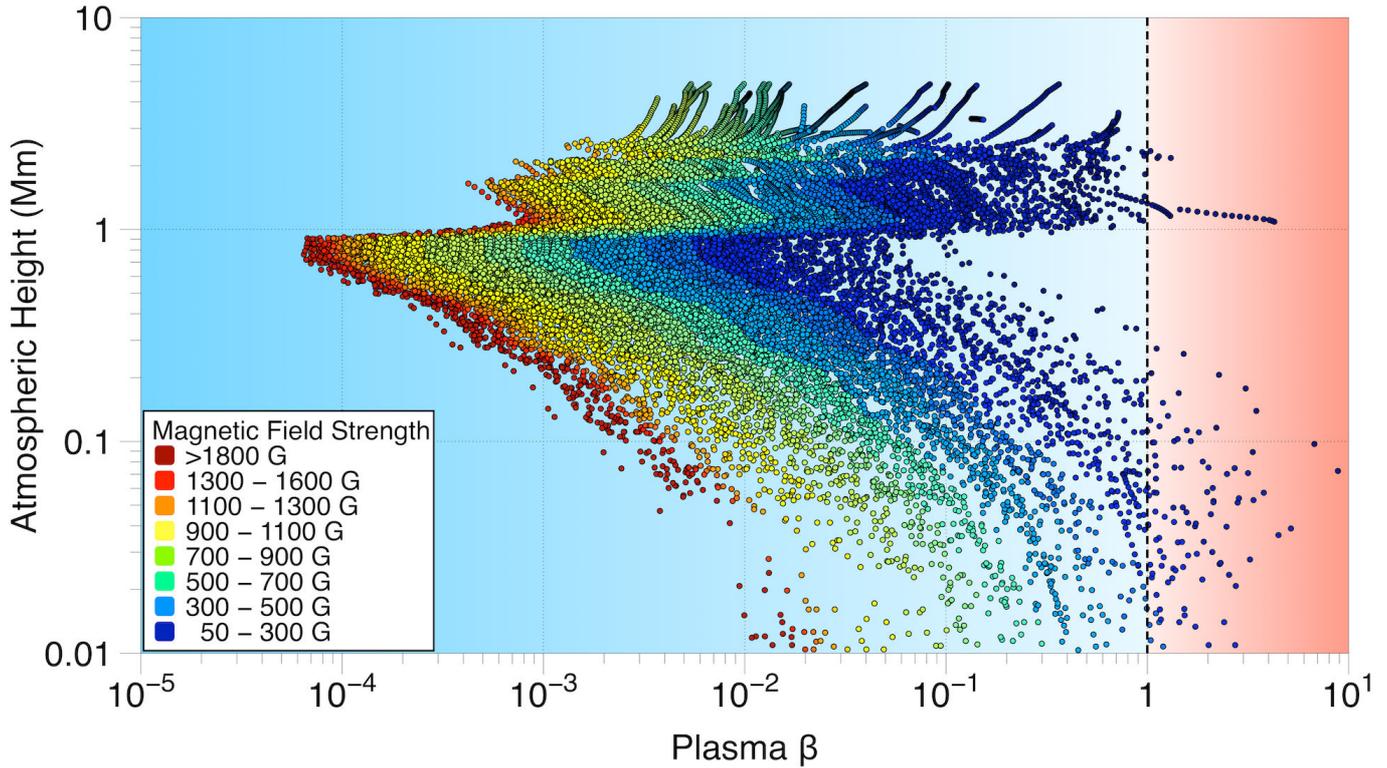

*Figure 4: **The degree of force-freeness of active region NOAA~12146.** Calculations of the plasma-β from the outputs of the force-fitted magnetic field topologies with minimized misalignment angles (providing the magnetic pressure), and the plasma pressure inferred from a well-established sunspot model atmosphere[S17], plotted as a function of atmospheric height. The individual data points are colored based upon the local magnetic field strength, as described by the color table in the lower-left quadrant of the plot. As expected, the strong magnetic field concentrations synonymous with sunspot umbral cores are predominantly force-free (i.e., plasma-β ≤ 1) across all atmospheric heights, hence validating the use of non-linear force-free magnetic field extrapolation codes for active region NOAA 12146.*

With no large-scale evolution of the umbral magnetic field confirmed, the application of a non-linear force-free extrapolation code to photospheric vector magnetogram data provides the most robust and accurate derivation of the stratified magnetic field geometry currently available[S30]. Of particular importance is the fact that the field exists within locations where the plasma-β < 1, which denotes an atmospheric region where the Lorentz force is negligible in comparison to the magnetic pressure, hence allowing the field to be described as *force-free*[S10,S31-S33]. A potential limitation of the methods applied here revolves around the use of the photospheric vector magnetogram data provided by SDO/HMI, where not all magnetic fields can be considered force-free[S34-S37]. The application of non-linear force-free field extrapolation codes to regions where the embedded field is not force-free has the potential to introduce mis-alignments between the extrapolated and real vector fields in the upper solar atmosphere[S38]. Fig. S4 indicates that the vast majority of the observed magnetic fields are force-free in the photosphere, a consequence of the high magnetic field strengths associated with the umbral core. However, the plasma-β values assigned to the outer sunspot magnetic fields, which are locations populated by lower magnetic field strength concentrations, are likely to be under-estimated due to the use of an umbral core density model[S17] in the calculations. For these locations, it is probable that the field lines are in fact β ≥ 1 in the lower atmosphere, consistent with previous findings[S39]. As a result, it is important to constrain the accuracy of the force-free extrapolations in light of the lower boundary conditions. This is achieved through the comparison of non-linear force-free field extrapolations with identified and traced features in high-resolution complementary chromospheric images (Fig. S3). Such comparisons reveal a three-dimensional mis-alignment between the observed chromospheric structures and the extrapolated field of ~10.7°, which is consistent with previous evaluations. A magnetic field mis-alignment of this magnitude further strengthens the conclusions of this work, since it precludes the possibility that the inclined magnetic fields near the outer boundary of the chromospheric umbra (which are necessary for the formation of Alfvén shocks; see Section 4) are mis-construed vertical fields.

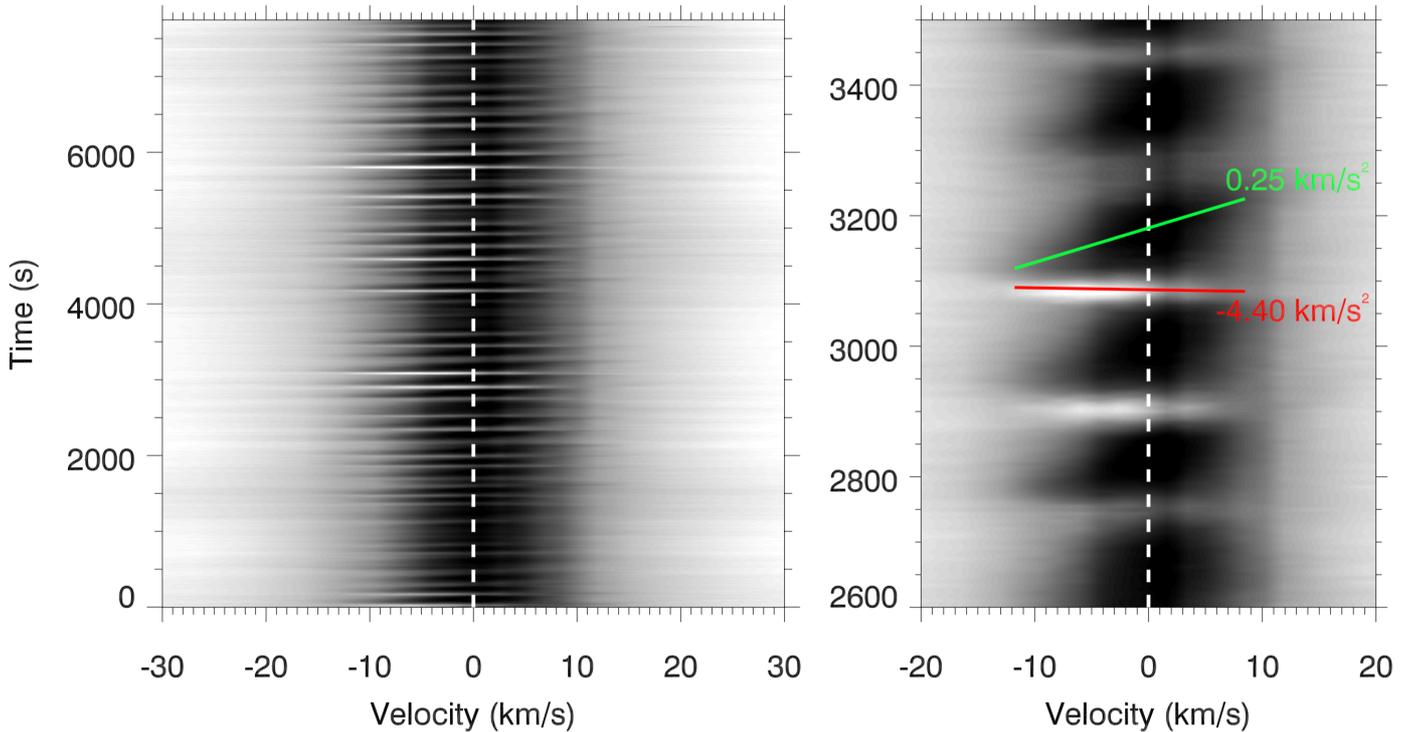

*Figure 5:* **The spectral signatures of umbral shock features.** *A velocity--time graph extracted from IBIS observations of the sunspot umbral core. Here, the horizontal axis has been converted from IBIS wavelengths into line-of-sight velocity space through the Doppler wavelength shifts away from the core of the chromospheric Ca II 8542 Å absorption line profiles. The brightnesses displayed correlate with the intensities of the corresponding Ca II 8542 Å spectral profiles. The left panel displays velocities of $\pm 30$ km s$^{-1}$ (or $\pm 0.85$ Å), corresponding to the full duration of the time series (135 minutes or 1336 IBIS scans), while the right panel zooms in to a smaller sub-set for a closer examination of the associated signatures. The red and green lines in the right panel denote the accelerations associated with the rising (blue-shifted) and falling (red-shifted) plasma, respectively.*

The small uncertainties associated with the extrapolated magnetic field mis-alignment angles will influence the plasma-β contours derived in Section 9 and depicted in Figs.~1 \& S9. However, small-scale fluctuations in the angle (inclination and/or azimuthal) of the embedded magnetic fields will not affect the conclusions surrounding the mode conversion discussed in Section 4, since the lateral expansion of the field lines will still introduce a region where coupling between magneto-acoustic and Alfvén modes will be prevalent. Nevertheless, it is clear that rectifying any magnetic field ambiguities, no matter how small, will enable future studies to better constrain the underlying plasma conditions responsible for Alfvén shock formation. Multiple avenues have been proposed to improve the accuracy of magnetic field extrapolations, including the integration of magneto-static photospheric models[S40,S41] with non-linear force-free field extrapolations, the modification and optimization of the base underlying vector magnetic field[S42,S43], and in future utilizing chromospheric magnetograms from full-Stokes measurements of magnetically sensitive lines (e.g., He I 10830 Å)[S44,S45] as lower boundary conditions to produce chromospheric and coronal extrapolations of unparalleled accuracy[S46]. In future years, a combination of the above proposed methods is likely to more robustly constrain the vector magnetic field in the solar chromosphere; not just in the umbra, but also in the surrounding quiet Sun where the magnetic pressure is substantially weaker.

## 3. Identifying Shock Events

In order to identify the locations of plasma shocks in the umbra, robust techniques were applied to minimize the number of false detections. These techniques are applicable to all sunspots, and so are of use to future investigations into shock phenomena. The first step was to isolate the chromospheric umbra from the penumbra, since the less magnetic penumbra hosts additional fluctuations that we do not wish to include in the present study, such as penumbral jets[S22] and running penumbral waves[S47,S48]. Simple intensity thresholding at each time step is impractical as it will exclude any bright flashing pixels contained within the sunspot, since these pixel intensities would be above the cut-off value used to contour the umbral perimeter. Instead, a time-average image was created for the entire Ca II 8542 Å core image sequence, which due to the lack of any change in sunspot structuring, produces an image with a high contrast ratio between the dark umbra and its immediate surroundings. This image was then manually thresholded to produce an accurate map defining the umbra/penumbra boundary, which was subsequently employed as a binary map[S49] to extract purely umbral time series for study. Furthermore, manual verifications were made to ensure that no bright intensity fluctuations were present in the pixels located at the umbra/penumbra interface throughout the time series. This ensured that all contributions from the penumbra (including dynamic penumbral events) were excluded from subsequent analysis.

Following the extraction of purely umbral time series, shocks were identified within the observations using running mean subtracted images of the line core[S51,S52]. Subtracting a running mean from the data allows for greater contrast between the flashes and the background umbra by removing any transient brightenings present for an interval longer than the flashes occupy, such as umbral dots[S50]. The images taken at the rest wavelength of 8542.12 Å were utilized, and the subtracted mean was calculated for each time step over ±25 images, corresponding to an approximate 5-minute window. The only exceptions were the first and last 25 images, where this was impossible. For these images, we instead subtracted the mean of the first and last 50 images, respectively, in line with previous studies[S52]. This ensures that only impulsive brightenings in the line core are considered for further study (Fig. S6).

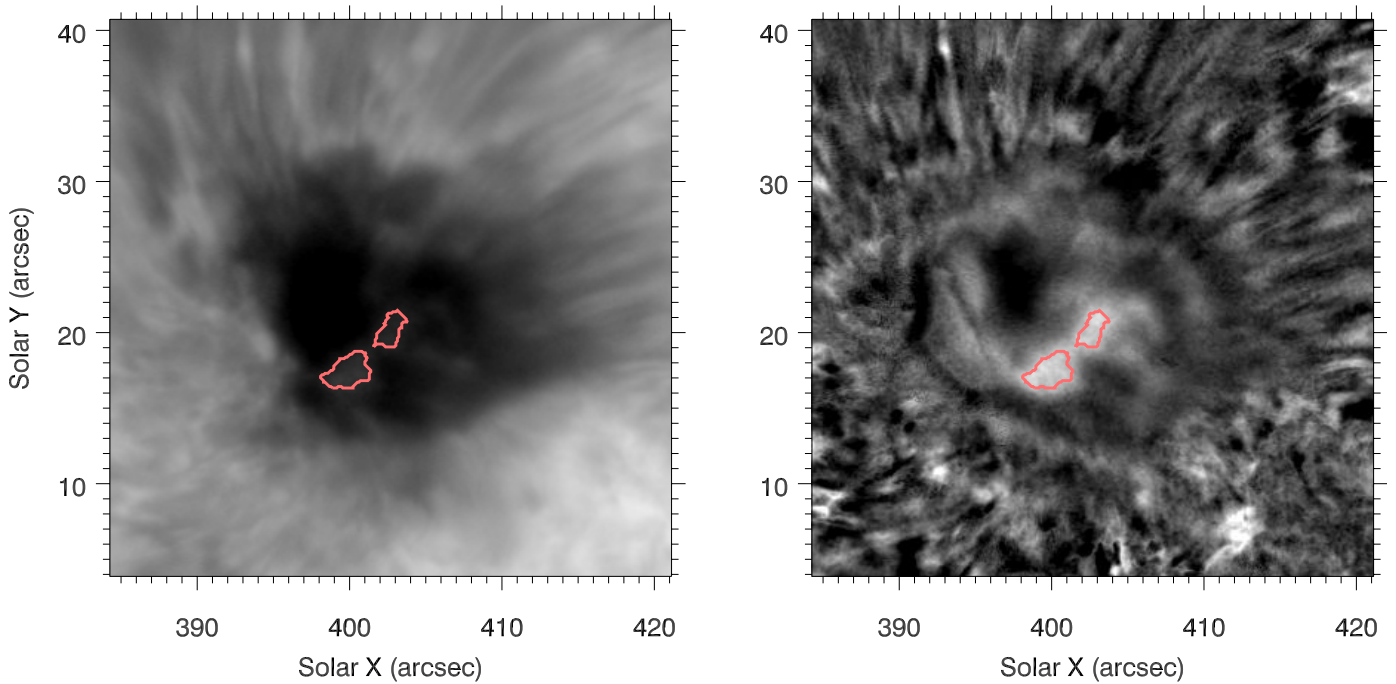

*Figure 6: **A visualization of statistically detected umbral shock features.** Images centered on the sunspot umbra, detailing the identification of shock events. The left panel is a line core image (taken at 8542.12 Å) and the right panel is the corresponding running mean subtracted image. Contoured in red is the thresholded region identified as shocks, which is much more apparent in the running mean subtracted image due to the thorough and robust background intensity renormalization.*

An advantage of running mean subtracted images is the normalization of umbral intensities through the removal of inherent (spatial) intensity gradients from the centre of the umbra outwards. With a normalized intensity, devoid of the natural umbral intensity structuring[S53], it is possible to accurately detect UFs throughout the umbra by means of intensity thresholding. As the mean-subtracted maps have an average value of zero, flash pixels were visually assessed to be any umbral pixel in the mean subtracted maps exhibiting an intensity excursion exceeding $2.7\sigma$ above zero, where $\sigma$ is the standard deviation of the mean subtracted map. The selection of a high threshold value ensures that no magneto-acoustic wave signals are included, as their intensity perturbations of ~10%[S54] fall far below the threshold set for shocks. A high threshold also mitigates any small-scale effects of photon noise present in the data. The manual identification of a suitable running mean subtracted intensity threshold also allowed for high number statistics, since arbitrarily selecting a larger value would naturally exclude weaker shock signals from being included in the study. Using this threshold, 554,792 UF pixels were detected within the 135-minute image sequence, allowing a statistical study of UF properties to be undertaken with unprecedented number statistics.

The criteria applied here ensure that no penumbral signals, long-term umbral brightenings, or linear magneto-acoustic waves can be detected. Therefore, the only signals present in this study are non-linear shock fronts in the impulsive phase of heating where they exhibit clear intensity spikes of at least 2.7σ above the umbral mean.

## 4. Alfvén Shocks

Alfvén waves are purely incompressible in the linear regime. Indeed, circularly polarized non-linear Alfvén waves remain incompressible as a result of the strict rotation of the magnetic field vector around the direction of propagation[S55]. However, linearly or elliptically polarized Alfvén waves, which naturally possess fluctuating magnetic field components, lead to density perturbations resulting from the induced non-linear ponderomotive forces. The non-linear dynamics of finite-amplitude Alfvén waves propagating along a straight unperturbed magnetic field ($B_z$) is governed by the derivative non-linear Schrödinger equation[S56],

$$\frac{\partial b}{\partial \tau} + \frac{1}{4(1-\beta)}\frac{\partial}{\partial z}(|b|^2 b) \mp i\frac{c_A^2}{2\Omega_i}\frac{\partial^2 b}{\partial z^2} = 0 \ , \quad (1)$$

where $b = B_\perp / B_z$ is the relative amplitude of transverse magnetic field perturbations, $\tau = (B_z / B_\perp)^2 t$ is the extended time, $\Omega_i$ is the ion cyclotron frequency, and $\beta \propto c_s^2/c_A^2$ is the ratio of the plasma pressure to the magnetic pressure, where $c_s$ and $c_A$ are the local sound and Alfvén speeds, respectively.

When the wave frequency is comparable to the ion-cyclotron frequency, the non-linear steepening (the second term in Equation 1) can be balanced by the wave dispersion (the third term in Equation 1), producing an Alfvén soliton solution[S57]. The plasma conditions necessary for the soliton solution are found, for example, in the dispersed solar wind. However, in the lower solar atmosphere, the ion-cyclotron frequency approaches $10^5 - 10^6$ Hz[S58], which negates wave dispersion and collapses the equation describing non-linear Alfvén waves to,

$$\frac{\partial b}{\partial \tau} + \frac{1}{4(1-\beta)}\frac{\partial}{\partial z}(|b|^2 b) = 0 \ , \quad (2)$$

highlighting the non-linear steepening of Alfvén waves[S59]. When the Alfvén speed diverges from the sound speed (i.e., $\beta \neq 1$), finite-amplitude Alfvén waves become capable of steepening into shocks[S60]. This steepening occurs over a wave period when the velocity amplitude becomes comparable to the Alfvén speed (i.e., $b \sim 1$).

Preferential conditions for Alfvén shock formation are found near the edge of an expanding magnetic flux tube, where the magnetic field lines become significantly inclined. Here, the magnetic field strength is continuously decreasing due to the lateral tube expansion, while the plasma density stays almost homogeneous, which leads to a negative Alfvén speed gradient. Alfvén wave energy flux in an expanding magnetic field is $\sim \sqrt{\rho}v^2 B_s A$, where $B_s$ is the magnetic field strength along a particular field line, $s$, and $A$ is the cross-sectional area normal to the field line. Continuity of the energy flux demands that the wave velocity amplitude, $v$, remains unchanged due to the magnetic flux conservation, thus ensuring the term $B_s A$ remains constant. However, the ratio of the velocity amplitude to the Alfvén speed can increase and approach unity ($b \sim 1$) as a result of the negative Alfvén speed gradient, which leads to the rapid steepening of Alfvén waves into shocks[S61].

The dissipation of Alfvén shocks, and the resulting plasma heating, remains an outstanding problem in solar physics. In the simplest case of a fully ionized plasma, Alfvén shocks are non-dissipative, since dissipation effects (e.g., viscosity) are neglected by definition. On the other hand, ion-neutral collisions in partially ionized chromospheric plasmas are likely to assist the dissipation of Alfvén wave energy into heat[S62,S63], though the modeling of such processes for Alfvén shocks is in its infancy.

The involvement of Alfvén waves in forming shock fronts is not limited to shocks formed from pure propagating Alfvén modes. When Alfvén waves propagate through a region with $\beta \sim 1$, then Equation 2 has a discontinuity, which allows for the resonant coupling of Alfvén and sound waves. Density perturbations associated with the ponderomotive force then begin to propagate as sound waves[S64,S65], which may steepen into magneto-acoustic shocks. The amplitudes of density perturbations are of the second order with regards to Alfvén wave amplitudes. There is, therefore, two defined scenarios for the formation of shocks in the outer umbra that both involve the presence of Alfvén waves: (1) Alfvén waves that directly steepen into shock fronts, and (2) sound waves that resonantly couple into Alfvén waves forming magneto-acoustic UFs in the outer umbra, both of which are visualized in Fig.~3 of the main text. The percentage of outer umbral shocks that are associated with each mechanism (Alfvén wave dissipation or resonantly coupled magneto-acoustic shocks) can be estimated through the investigation of the Doppler velocity signatures related to the shock fronts, as described in the section below.

**5. Doppler Velocity Measurement**

When considering Ca II 8542 Å spectral profiles, the line-of-sight (LOS) Doppler velocity is calculated by measuring the wavelength shift of the line core from its well-defined rest wavelength. Normally, the line core is defined by fitting a Gaussian to the absorption profile. For shock phenomena this is not possible, since from inspection of Fig. S1, the core emission of shocks adds complexity to the overall profile shape. Without proper consideration of this emission, misleading LOS velocity values can be derived for the shock plasma[S66].

To remedy this issue, resultant emission profiles were created by subtracting the time averaged profile at each shock location from the shock profiles evident at that moment in time (black line in Fig. S1). The resultant profiles take the form of emission profiles representing the spectral characteristics of the shocked plasma at that particular instant. Subsequently, the LOS velocities are calculated using cumulative distribution functions of intensity against wavelength[S67], where the line core is defined as the wavelength where 50% of emission has occurred. The robustness of this method results from no underlying assumptions being made regarding the shape of the spectral profile, allowing for the determination of the true line core for shocked profiles exhibiting both blue- and red-shifted emission. Despite a lack of detailed modeling verifying the accuracy of quartile analysis[S67], it accurately portrays the spot center UFs as only exhibiting blue shifted emission, consistent with the established picture of UF evolution, and indeed in agreement with Fig. S5. As such, the cumulative distribution function method can be accurately applied to the shocks identified in this study in order to determine their intrinsic LOS velocities.

Figs.~2 \& S5 show that the majority of LOS plasma velocities are blue-shifted and display characteristic `sawtooth' spectral profiles throughout their temporal evolution. This is consistent with the established morphology of traditional UFs, whereby their initial motion is upwards and in the direction of wave propagation, before the plasma cools and sinks to it's equilibrium position[S51,S68,S69]. Any appearance of red-shifted velocities during impulsive shock formation must be a signature of

Alfvén shocks, since UFs only exhibit red shifts when the plasma cools, thus the weaker emission associated with this stage of evolution will prevent the Doppler signatures from being detected by the thresholding criteria. Furthermore, all optical depth and derived temperature values in this study are decoupled from the LOS velocity, ensuring that the observed red shifts are not a result of changes in opacity[S70] (see Section 6 below). Approximately 35% of the velocities found in the outer umbral shock population in Fig.~2 of the main text are positive. Since the velocity amplitudes of the Alfvén waves will be comprised of an equal number of blue- and red-shifted components, this provides an estimation of ~70% (i.e., 35% red-shifted combined with their 35% blue-shifted plasma counterparts) resulting from the presence of Alfvén waves. Therefore, at least ~70% of the shocks found within the outer umbral region are caused by the direct steepening of Alfvén waves.

## 6. Temperature Inference Method

The thermal structuring of the imaged region was derived using the CAlcium Inversion using a Spectral Archive[S71] (CAISAR) code developed specifically for the Ca II intensity profiles obtained from IBIS observations. This method assumes a local thermodynamic equilibrium (LTE) atmosphere and derives the thermal properties through the comparison of the observed profiles to a synthesized archive encompassing a large variety of temperature perturbations[S72-S74]. A first-order correction to a non-LTE environment has been established and verified[S75], allowing for accurate chromospheric temperature differentials to be derived from a corrected LTE framework.

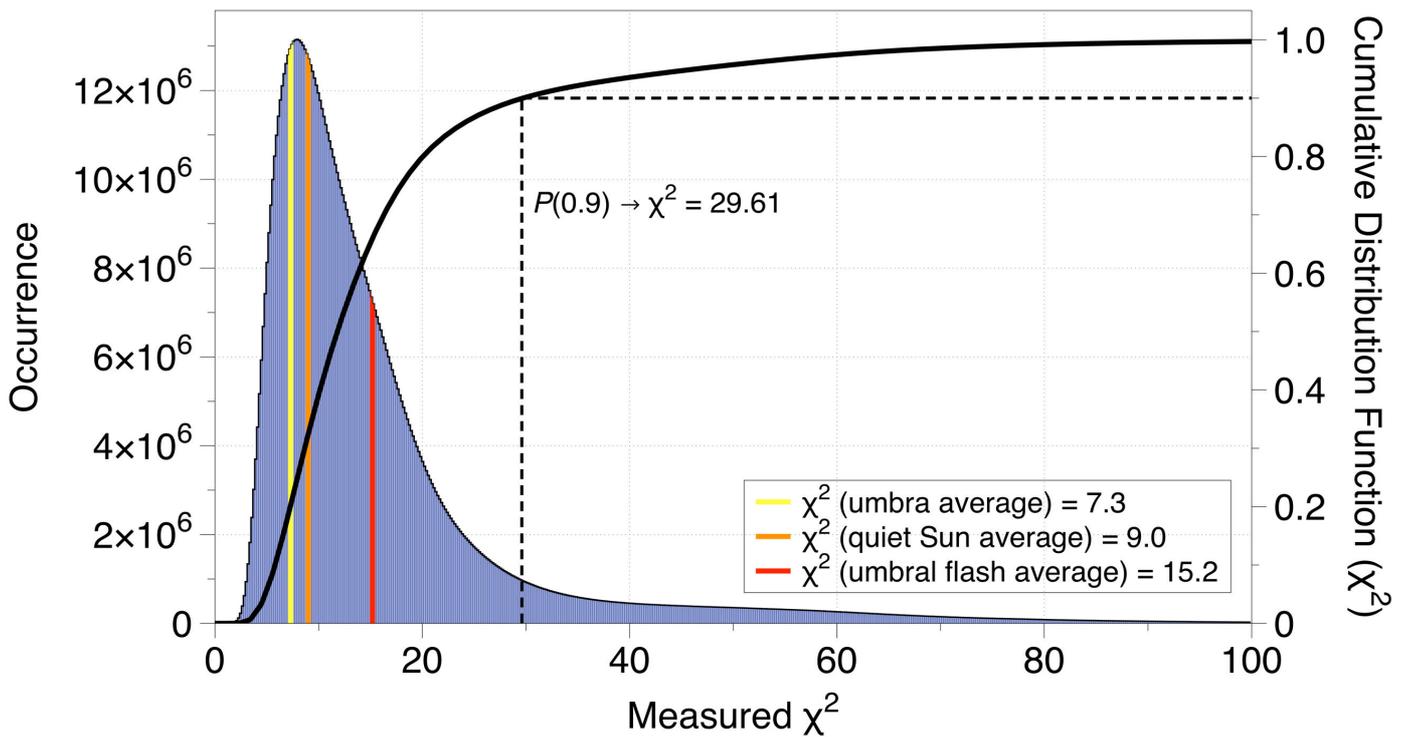

*Figure 7: $\chi^2$ **statistics of the CAISAR temperature inversion outputs.** A histogram of the measured $\chi^2$ values (blue bars), over-plotted with the cumulative distribution function (black line), which corresponds to a $\chi^2$ distribution demonstrating 21 degrees of freedom. Vertical yellow, orange and red lines indicate the average umbral, quiet Sun and flash $\chi^2$ values, respectively. A vertical dashed black line links a cumulative distribution function probability of 0.9 to its expected $\chi^2$ value, in this case $P(0.9) \rightarrow \chi^2 = 29.61$. The average umbral, quiet Sun and flash $\chi^2$ measurements are all below $P(0.66)$, indicating robust spectral fitments. Values beyond $\chi^2 = 29.61$ would indicate relatively poor spectral fits.*

The efficiency of CAISAR allowed for the entire dataset, approximately $10^9$ individual Ca II 8542 Å profiles, to be inverted in approximately 5 months on a 2.90 GHz Intel Xeon processing cluster incorporating 32 CPU cores. The output of the code for each of the 1336 spectral scans is a three-dimensional cube, comprising of the observational spatial dimensions and 75 points in optical depth ($-6 < log\ (\tau_{500nm}) < 0$), where $\tau_{500nm}$ is the optical depth corresponding to a wavelength of 500 nm. A $\chi^2$ measurement of fit quality was obtained for each inverted profile, which revealed that reliable fits were achieved across the entire field-of-view. The $\chi^2$ values were computed for each spectral fit according to the relation[S76],

$$\chi^2 = \sum_{\lambda=1}^{N} \frac{(I_\lambda^{obs} - I_\lambda^{syn})^2}{I_\lambda^{syn}},$$

where $I_\lambda^{obs}$ and $I_\lambda^{syn}$ are the measured and synthetic intensities, respectively, at each wavelength, $\lambda$, while N represents the number of sampled wavelength positions (N = 27; Fig. S1). A histogram of all measured $\chi^2$ values (Fig. S7) reveals a clear Poisson-like distribution that is characteristic of expected discrete variable comparisons[S77]. The cumulative distribution function of the $\chi^2$ histogram is consistent with a distribution that has 21 degrees of freedom. This agrees with the expected behavior for the spectral fits. Five fitting parameters, n=5, were used in the generation of the LTE archive, incorporating global offsets in temperature (1 parameter), the additions of straight lines of arbitrary slope to the temperature stratifications (1 parameter), and the addition of variable Gaussians with free parameters in their widths, amplitudes and locations in optical depth (3 parameters)[S72]. The histogram of the $\chi^2$ values should therefore correspond to a distribution with total degrees of freedom, $\nu$, equal to $\nu$ = N - n - 1 = 27 - 5 - 1 = 21.

To further quantify the $\chi^2$ measurements, a quiescent region away from the sunspot was compared with the umbra and flash profiles to assess the quality of fit. For the quiet Sun region, $\chi^2 \approx 9.0$, with the umbra and flash locations exhibiting $\chi^2 \approx 7.3$ and $\chi^2 \approx 15.2$, respectively. Given that the LOS velocities associated with shocks produce a Doppler-shifted spectral profile (see, e.g., Fig. S1), higher $\chi^2$ values for these locations are naturally expected, though overall the measurements show that the inversion code produces accurate thermal fits irrespective of the underlying solar feature. This is particularly apparent when considering that the umbral, quiet Sun and flash $\chi^2$ fit values demonstrate cumulative distribution function probabilities of 0.20, 0.32 and 0.66, respectively, showing that the typical spectral fits are accurately constrained by the CAISAR inversion code (see the vertical yellow, orange and red lines in Fig. S7).

## 7. Thermal Atmospheric Response to Shocks

In order to study the general thermal perturbations caused by the UFs and Alfvén shocks present in the data, the thermal response profiles, as a function of optical depth, were averaged across all detected UF/Alfvén shock pixels. The resulting shock thermal profiles are seen in the left panels of Fig. S8, alongside quiescent (i.e., non-shocking) umbral temperature profiles corresponding to the same spatial locations. The deviations of the percentage shock temperatures, with respect to the local quiescent umbral averages, are seen in the right panels of Fig. S8, which is useful for investigating the preferential locations of temperature perturbations as a function of optical depth. The shock profiles conform closely to the average umbral temperatures at higher optical depths (i.e., $log\ (\tau_{500nm}) \rightarrow 0$), equating to minimal temperature fluctuations at lower, photospheric heights. The first temperature variations associated with shocked plasma occur at around the location of the umbral temperature minimum, corresponding to $log\ (\tau_{500nm}) \sim -2$, or an atmospheric height of approximately 250 km above the photosphere. A significant temperature discrepancy for UFs begins to appear at $log\ (\tau_{500nm}) \sim -3$, with a 5% increase in temperature from the background quiet umbra. Again, at optical depths of $log\ (\tau_{500nm}) \sim -3$, the Alfvén shock temperatures begin to deviate from the local background temperature structuring, albeit with significantly smaller amplitudes on the order of a fractional percentage increase. As the optical depth continues to decrease, the temperatures for both UFs and Alfvén shocks continue to separate from their local background mean, with the largest temperature differentials occurring at $log\ (\tau_{500nm})$ = -4.9 (~16% rise; UFs) and $log\ (\tau_{500nm})$ = -5.1 (~6% rise; Alfvén shocks). From Fig. S8, it is clear that UFs and Alfvén shocks occur at similar optical depths (predominantly within the interval -5.3 < $log\ (\tau_{500nm})$ < -4.6), yet Alfvén shocks only produce around one third of the fractional temperature enhancement through wave dissipation that UFs provide.

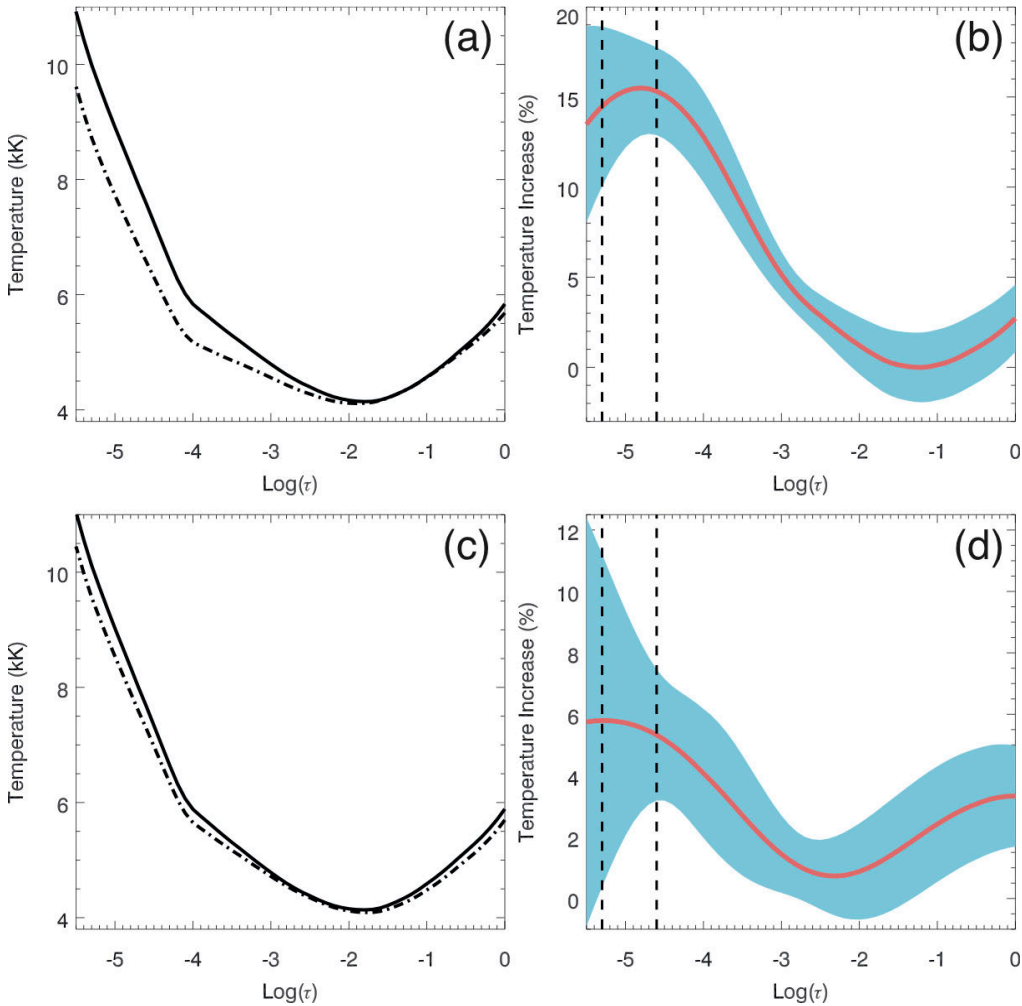

Figure 8: **The thermal response of the umbral atmosphere to plasma shocks.** *The left panels plot the averaged UF (panels **a** and **b** and Alfvén shock (panels **c** and **d**) temperature profiles in bold with respect to the logarithm of optical depth, whereas the dot-dashed lines represent the average umbral temperature profiles manifesting in their respective locations. The right panels illustrate the percentage temperature increases of the UFs (top) and Alfvén shocks (bottom) with respect to their local quiescent umbral temperatures as a function of optical depth (red line), with the standard error at each optical depth visualized as a blue shaded region. The vertical dashed lines represent the range of optical depths (-5.3< log ($\tau_{500nm}$) <-4.6) containing a temperature peak for both UFs and Alfvén shock phenomena.*

The calculated temperature differentials arising between traditional UFs and Alfvén shocks can be attributed to the energetics of the magneto-acoustic waves driving the shock formation. The inherent gravity of the Sun ensures that only short wavelength, more energetic magneto-acoustic waves can propagate along vertical magnetic fields[S78], whereas the more inclined fields present in the region corresponding to Alfvén shocks allow for the propagation of longer wavelength, less energetic magneto-acoustic waves from the photosphere into the chromosphere[S48]. Thus, the magneto-acoustic waves that mode convert into Alfvén waves, and subsequently steepen into shocks, provide less energy to the overall shock front, which leads to smaller temperature perturbations than the more energetic magneto-acoustic shocks present towards the center of the umbra.

In comparison to the only other example of generalized temperature values derived for UFs[S79], we uncover similar thermal morphologies as a function of optical depth. However, a one-to-one comparison is limited by the small UF sample contained within the former study and the different temperature inversion routines employed. The sample size of spectral profiles involved in the present analysis makes Fig. S8 the most comprehensive insight into UF temperature fluctuations to date.

From inspection of Fig. S8, a proportion of the relative temperature fluctuations associated with Alfvén shocks display a reduction in the plasma temperature during the onset of shock dissipation. Here, the resulting spectral signatures are dominated by the background umbral atmosphere, hence providing the first imaging evidence of the two-component shock atmosphere model describing the superposition of quiescent umbral and shocked plasma spectral profiles, which have previously only been detected in magnetically sensitive Stokes-V signals[S80-S82].

The average temperature fluctuations intrinsic to UFs and Alfvén shocks are a valuable tool for defining optical depths of interest for further study. Due to the inverted temperature values being a function of optical depth, it is impractical and imprecise to select a single optical depth for subsequent study, especially since UF and Alfvén shock temperature enhancements span a broad range of optical depths (Fig. S8). Therefore, a range of $log\ (\tau_{500nm})$ values were selected to ensure adequate temperature fluctuation coverage, spanning a range of $-5.3 < log\ (\tau_{500nm}) < -4.6$, which encompasses the optical depths where the largest percentage deviations in temperatures from the quiescent local umbral means are seen for both UFs ($log\ (\tau_{500nm}) = -4.9$) and Alfvén shocks ($log\ (\tau_{500nm}) = -5.1$). Temperature fluctuations extracted from within this range of optical depth values are employed for Figs. 2 & 4 in the main text.

Variations in the geometric formation height of shocks can also be seen through estimates made from the observed data in this study. Applying the optical depths where temperature enhancements begin to manifest in Fig. S8 ($log\ (\tau_{500nm}) \sim -2$) to sunspot model atmospheres[S17] produces an estimated column height beginning at ~250 km, and stretching to ~1100 km where the peak percentage temperature perturbations from the mean occur ($log\ (\tau_{500nm}) \sim -4.9$). This can be favorably compared to recent modeling work[S83,S84], which saw evidence of temperature enhancements manifesting at heights of approximately 700 km and 400 km above the solar surface. Furthermore, this is also consistent with the IBIS contribution function for the Ca II 8542 Å line core[S29] where significant shock activity is observed. There is, therefore, a consensus on the general region where these shocks are forming. However, the exact heights observed cannot be deduced until the field of 3D MHD modeling of sunspots advances to a stage where the observational parameters can be directly compared.

## 8. Establishing the Local Plasma Density

We employed the Non-LTE Inversion Code using the Lorien Engine[S85] (NICOLE) to infer the approximate physical conditions in the atmosphere at each spatial location. This technique, applied to a selection of 200 shocking umbral pixels captured in the Ca II 8542 Å infrared line, has been tested extensively against the most realistic chromospheric simulations[S86]. The non-LTE inversions with NICOLE require several seconds per profile, and thus could not be run over all of the $10^9$ sampled Ca II 8542 Å spectra in a reasonable time. The model atom considers 5 bound levels plus the Ca II continuum, treated in complete angle and frequency redistribution. Inside each pixel, the atmosphere is considered homogeneous and infinite in the horizontal direction.

The synthetic profiles are computed on a wavelength grid much finer than the observations, in order to allow for a proper treatment of frequency redistribution and spectral broadening due to the instrumental profile. However, only those points at the observed wavelengths are considered in the comparison to the observations. The rest of the synthetic profile samples have zero weight in the $\chi^2$ merit function.

In order to reproduce the rich structure of the chromospheric Ca II 8542 Å line cores in the observations, the code is run with a large number of nodes (most notably 11 in temperature and 8 in line-of-sight velocity). Instabilities are avoided by using a regularization scheme that penalizes any departures from smooth solutions. The penalization is included as an extra term in the merit function, weighted by a parameter that controls the regularization strength. This sets the balance between smoothness and the goodness-of-fit. In practice, the parameter is chosen so that the regularization term always remains at least an order-of-magnitude smaller than the total $\chi^2$. The resulting average sunspot umbral density, $\rho = 3.57 \pm 1.30 \times 10^{-7}$ kg m$^{-3}$, corresponding to the peak of the Ca II 8542 Å contribution function, is consistent with the region where maximum shock-induced temperature perturbations are found ($-5.3 < log\ (\tau_{500nm}) < -4.6$), agreeing with the empirical model atmospheres[S17] used to establish the preferential locations for efficient mode-conversion of magneto-acoustic waves.

## 9. Calculation of the Plasma Beta

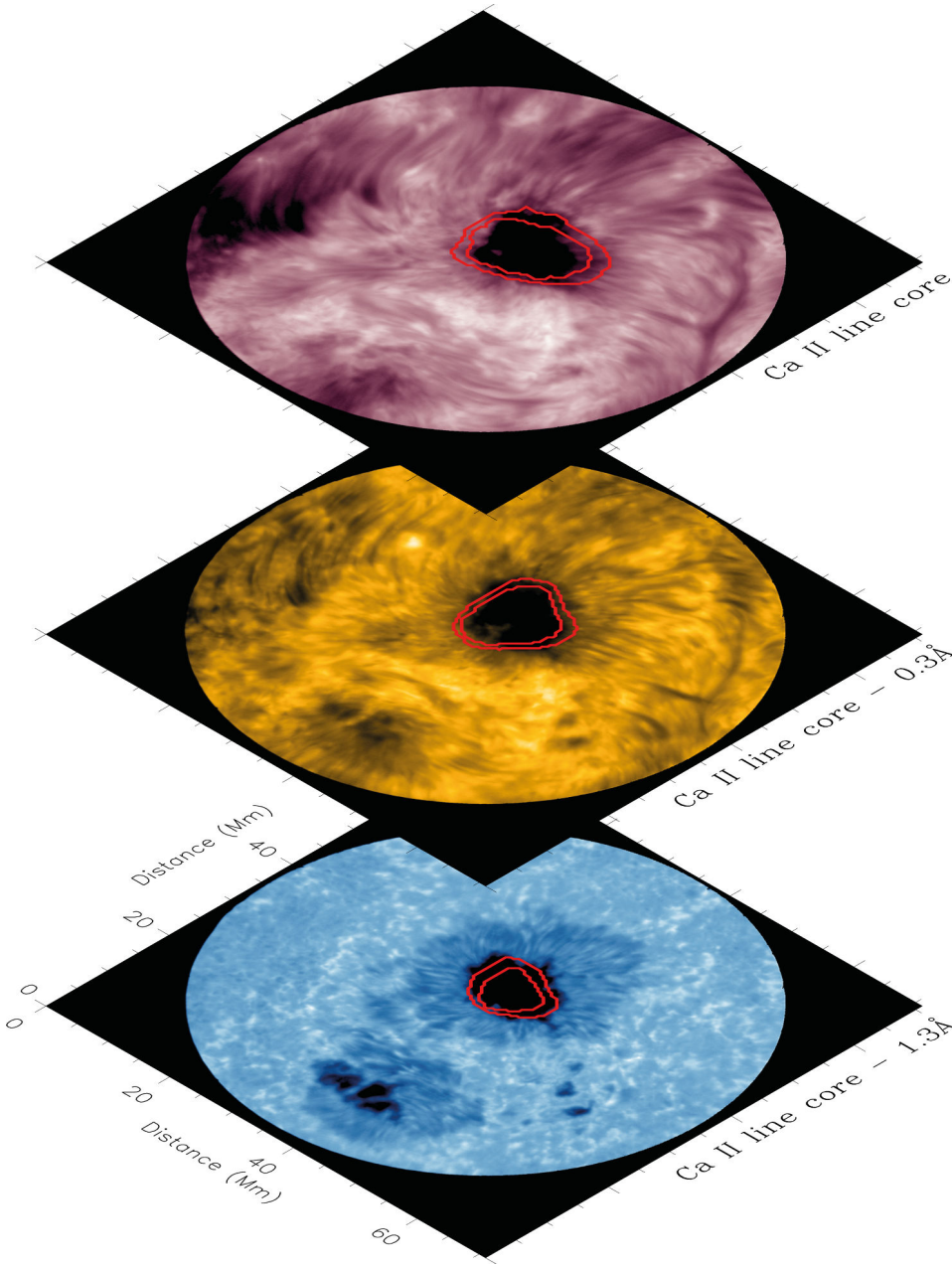

*Figure 9: **Locations of mode conversion at various atmospheric heights.** A stack of IBIS Ca II 8542 Å spectral images that, from the bottom upwards, represent the photosphere, lower chromosphere and upper chromosphere, respectively. Similarly, the mode conversion regions, which are encapsulated within red contours, have been calculated for optical depths of log ($\tau_{500nm}$) = -2.0 (~ 250 km; where shocks are first detectable), log ($\tau_{500nm}$) = -3.0 (~ 450 km; more pronounced shock features) and log ($\tau_{500nm}$) = -5.0 (~ 1000 km; greatest temperature enhancements). It can be seen that the mode conversion region expands laterally with atmospheric height, a consequence of the expansion of the local magnetic fields and consistent with the trends depicted in Fig. S4.*

With the signatures of Alfvén shocks clearly present in the data, consideration turns to the origin of the Alfvén waves forming these shocks. With recent modeling indicating that the generation of Alfvén waves can be inhibited in strongly magnetic photospheric regions[S87], other mechanisms must be considered to generate the energetic waves capable of forming the observed shocks.

The process of mode conversion allows magneto-acoustic waves to efficiently convert into Alfvén modes when their characteristic velocities, namely the sound speed and the Alfvén velocity, are equivalent[S65,S88]. For observational data with non-infinite resolution, the locations where the characteristic velocities are equivalent can be closely mapped to the plasma-β = 1 region[S89], which can be easily derived from the data products present for this study. The plasma-β is defined as the ratio between the plasma gas pressure and the pressure of the magnetic field, and can be derived from the temperature of the plasma (*T*) and the magnitude of the magnetic field ($B_{tot}$) following[S48],

$$\beta = \frac{2\mu_0 n_H T k_B}{B_{tot}^2},$$

where $\mu_0$ is the magnetic permeability, $n_H$ is the hydrogen number density and $k_B$ is the Boltzmann constant. A time-averaged thermal map (including associated uncertainties) was created and combined with the extrapolated chromospheric vector magnetic field maps to derive plasma-β ratios for all locations within the field-of-view corresponding to the atmospheric heights that begin to first demonstrate UF and Alfvén shock temperature enhancements from the background mean (*log* ($\tau_{500nm}$) ~ -2; 250 km; Fig. S8). It should be noted that the extrapolated magnetic fields, which define the magnitude of the magnetic pressure, are computed at the same spatial sampling as the SDO/HMI vector magnetograms (i.e., 0."5 or 362.5 km per pixel). As a consequence, when the plasma-β = 1 isocontours are traced on top of the higher resolution IBIS observations, a slight 'coarseness' is visible (see, e.g., Fig. S9). The density employed is extracted from a sunspot model atmosphere (model 'M')[S17], where $n_H \approx 6 \times 10^{16}$ cm$^{-3}$ represents the deepest layer of the solar atmosphere where UFs and Alfvén shocks are detected. A large component of the outer umbra has the plasma-β = 1 (see Fig. 1 in the main text), providing ideal locations for complex wave interactions.

The use of a semi-empirical density model will naturally add a level of uncertainty to the computed plasma-β maps. Employing a constant model 'M' density[S17] across the entire umbra is likely to under-estimate the real densities present towards the outer umbral boundary. As a consequence, the diameters of the calculated plasma-β = 1 contours are likely to be too large (Figs. 1 & S9), since an increase in the local density towards the outer umbral edge will act to boost the local gas pressure and hence more quickly establish a plasma-β = 1 environment at smaller umbral radii. However, this does not mitigate the implications surrounding the locations of the plasma-β = 1 contours, since they will remain within the umbra, thus continuing to allow for an extended region where the mode conversion of upwardly propagating magneto-acoustic umbral waves into Alfvén modes is possible. Of course, not all of the magneto-acoustic waves incident on the plasma-β = 1 region are expected to mode convert, since magneto-acoustic signals are observed to propagate beyond this layer and into the transition region and corona[S90,S91]. However, mode conversion in the plasma-β = 1 layer provides an efficient mechanism for the creation of Alfvén waves that can subsequently dissipate their energy through the process of shock heating[S92].

Pioneering work[S61] has revealed that Alfvén shocks are likely to manifest in regions displaying a large negative gradient in the local Alfvén speed. For the present dataset, we are able to map a two-dimensional representation of the local Alfvén speed, $c_A$, through the relation,

$$c_A = \frac{B_{tot}}{\sqrt{\mu_0 \rho}},$$

where $\rho$ is the local plasma density. The magnitude of the magnetic field strength, $B_{tot}$, is extracted from the magnetic field extrapolations synonymous with the atmospheric height where UFs and Alfvén shocks first begin to manifest (~250 km; just above the umbral temperature minimum). The plasma density, $\rho \sim 10^4$ kg m$^{-3}$, is consistent with the hydrogen number density used above for the deepest layer of the solar atmosphere where UFs and Alfvén shocks are detected[S17], thus remaining consistent with the umbral environment promoting the initiation of mode converted Alfvén waves at an atmospheric height of approximately 250 km. The density is kept constant throughout the field-of-view for the purposes of calculating a two-dimensional map of the local Alfvén speeds (Fig. S10). This density may be too low for the quiet Sun regions away from the umbra, thus overestimating the Alfvén speeds in these locations. However, since the magnetic field strengths in these quiet Sun locations are significantly reduced (when compared with the sunspot umbra), the resultant Alfvén speed approaches zero irrespective of the density chosen. Importantly, in the umbra, where the Alfvén shocks are occurring, this density value results in Alfvén speeds consistent with previously determined values employing seismological approaches[S93].

The computed Alfvén speeds are on the order of 22 km s$^{-1}$ near the umbral core where the magnetic field strengths are strongest, dropping to approximately 5 km s$^{-1}$ towards the outer regions of the penumbra (Fig. S10). The locations synonymous with the plasma-β =1 isocontours demonstrate

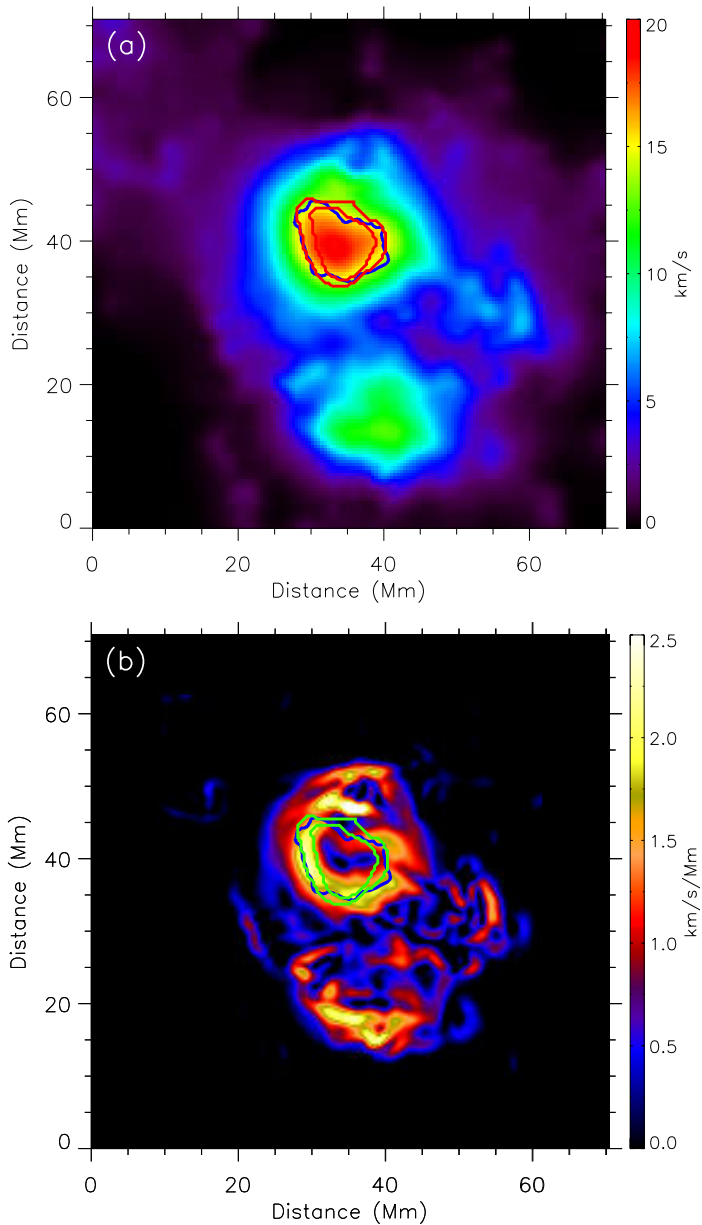

*Figure 10:* **Two-dimensional maps of the Alfvén speed structuring in the vicinity of the sunspot.**

*(a)* A two-dimensional representation of the Alfvén speed constructed for an atmospheric height of approximately 250 km (where magneto-acoustic-to-Alfvén wave mode conversion first manifests), where the colorbar indicates the local Alfvén speed in km s$^{-1}$. *(b)* The spatial derivative of the Alfvén speed, where the dark-through-bright color scheme indicates negligible-to-significant Alfvén speed gradients, as quantified by the colorbar in units of km s$^{-1}$ Mm$^{-1}$. Elevated negative Alfvén speed gradients are present within the confines of the plasma-$\beta$=1 isocontour (red and green lines in the left and right panels, respectively). The blue contour depicts the umbra--penumbra boundary established from the time-averaged IBIS spectral images.

Alfvén speeds consistent with typical chromospheric sound speeds, reiterating the validity of an equipartition layer existing within the outer umbral atmosphere. Importantly, encapsulated within the plasma-$\beta$ = 1 isocontours is a significant negative Alfvén speed gradient, changing from ~16 km s$^{-1}$ to ~10 km s$^{-1}$ over the spatial extent of only a few thousands of km. To spatially map the Alfvén speed gradients present in the data, a two-dimensional Savitzky-Golay[S94,S95] spatial derivative filter was constructed and convolved with the Alfvén speed map, providing a map of the associated Alfvén speed gradients (Fig. S10). The increase in the magnitude of the (negative) Alfvén speed gradient is co-spatial with the inner confines of the plasma-$\beta$ = 1 boundary. Thus, a significant negative Alfvén speed gradient, coupled with a co-spatial plasma-$\beta$ = 1 equipartition layer, provides the idealized environment for magnetoacoustic-to-Alfvén mode conversion and subsequent Alfvén shock formation. Through inspection of the shock-induced tangential Doppler velocities, we provide, for the first time, evidence of localized heating by Alfvén wave dissipation in the solar chromosphere.

## 10. Summary of Alfvén Wave Calculations

In this work, we establish evidence that, for the first time, details the mode conversion and dissipation of Alfvén waves in the solar chromosphere of a sunspot umbra. Initially, while examining traditional umbral flash signatures produced by active region NOAA 12146, a second population of shock signatures was identified near the periphery of the umbral/penumbral boundary. This new population demonstrated both blue *and* red-shifted Doppler velocities during the initial shock formation, with the velocities manifesting perpendicularly to the wave vector. These characteristics separate them from traditional umbral flashes that only display blue-shifted signatures along the direction of the wave vector. The signatures are the result of upwardly-propagating magneto-acoustic waves mode-converting into Alfvén waves, which are then subsequently able to form shock fronts (see Section 4).

Following the application of the CAISAR inversion routine on the Ca II 8542 Å spectral observations, temperature fluctuations caused by the shocking Alfvén waves are first detectable at *log* ($\tau_{500nm}$) ~ -2, corresponding to an atmospheric height of approximately 250 km above the umbral photosphere.

More pronounced fluctuations are visible at *log* ($\tau_{500nm}$) ~ -3 (or 450 km above the photosphere), with the largest temperature excursions (~6% rise) established for *log* ($\tau_{500nm}$) ~ -5, which relates to an atmospheric height of approximately 1000 km above the surface. Therefore, the column depth where Alfvén shocks readily form spans the atmospheric heights of approximately 250 – 1100 km. As a result, the mode conversion of magneto-acoustic waves into Alfvén modes must begin to occur close to the base of the shocking zone. In order to calculate the plasma-$\beta$ = 1 region, where mode conversion processes are most efficient, temperatures (from the CAISAR inversions at *log* ($\tau_{500nm}$) ~ -2), magnetic field strengths (from the non-linear force-free extrapolations at an atmospheric height of approximately 250 km) and a hydrogen number density of $n_H \approx$ 6 x $10^{16}$ cm$^{-3}$ (corresponding to $\rho$ ~ $10^4$ kg m$^{-3}$ from a well-established sunspot umbral model atmosphere at *log* ($\tau_{500nm}$) ~ -2) were selected for the purposes of calculating maps of the local magnetic and plasma pressure. The ratio between the two pressure maps provides a two-dimensional representation of the plasma-$\beta$ parameter, with the plasma-$\beta$ = 1 region contoured in Figs. S10 & 1 of the main text. It is within the plasma-$\beta$ = 1 contours (at an optical depth of *log* ($\tau_{500nm}$) ~ -2, or an atmospheric height of approximately 250 km) where mode-conversion of magneto-acoustic modes into Alfvén waves first begins to occur.

Once the magneto-acoustic waves mode-convert into Alfvén waves, they will continue to propagate upwards along the magnetic field lines that become increasingly more inclined towards the edge of the sunspot umbra. To accurately evaluate the Alfvén wave energy, $E_A$, at the point where the Alfvén waves first begin to shock, it is paramount to be able to identify the precise location where a plasma shock forms. This identification is performed through analysis and thresholding of the running mean subtracted IBIS time series (see Section 3). With the times and locations of initial Alfvén shocks extracted from the data, the corresponding Ca II 8542 Å spectra were used with the non-LTE NICOLE inversion code to deduce the plasma densities synonymous with the formation of Alfvén shocks, where a density of

$\rho$ = 3.57 ± 1.30 x $10^{-7}$ kg m$^{-3}$ is found. When compared with empirical sunspot model atmospheres, this density is consistent with the optical depths and atmospheric heights where peak Alfvén shock induced temperature perturbations are found (-5.3 < *log* ($\tau_{500nm}$) < -4.6, or 750 - 1100 km above the photosphere).

To subsequently calculate the Alfvén wave energy, the computed density ($\rho$ = 3.57 ± 1.30 x $10^{-7}$ kg m$^{-3}$) is combined with the measured velocity amplitude (v = 2.2 ± 0.6 km s$^{-1}$) and the Alfvén speed ($c_A$ = 6.2 ± 0.5 km s$^{-1}$; extracted from the non-linear force-free extrapolations at an atmospheric height of approximately 1000 km) synonymous with the optical depths (-5.3< *log* ($\tau_{500nm}$) <-4.6) demonstrating peak temperature enhancements. This provides an Alfvén wave energy estimate of $E_A$ = 10.7 ± 5.7 kW m$^{-2}$. This energy is substantially less than the 20 kW m$^{-2}$ observed for upwardly propagating magneto-acoustic waves generated in photospheric sunspot umbrae (see main text), which highlights that only a portion of the incident magneto-acoustic waves will couple into Alfvén waves (and subsequently shock) within the mode conversion layer.

Interestingly, a velocity amplitude, v = 2.2 ± 0.6 km s$^{-1}$, combined with a local Alfvén speed, $c_A$ = 6.2 ± 0.5 km s$^{-1}$, provides an Alfvén Mach number of *b* = 0.35 ± 0.10 (see Section 4). While this may initially suggest the presence of weakly non-linear effects, previous numerical simulations have demonstrated that the continuous driving of Alfvén waves (and subsequent shocks) can actually produce strong non-linear shocks at Alfvén Mach numbers of *b* ~ 0.20[S61]. This is a consequence of the plasma at the front of the subsequent wave trains impacting into the post-shocked material, thus enhancing the wave steepening and shock formation. Indeed, for the period, T ~ 180 s, of the continuous waveforms ultimately producing Alfvén shocks (see Fig. S5), it has been found[S61] that they can readily develop with velocity amplitudes on the order of 2.5 km s$^{-1}$, which is consistent with the velocity amplitudes established in the present study. Furthermore, when the period of the underlying wave motion, T ~ 180 s, is coupled with the Alfvén speed synonymous with the outer umbral regions, $c_A \approx 6.2$ km s$^{-1}$, it becomes possible to estimate the thickness, *L*, of the mode conversion region where Alfvén shocks may be initiated as L = T $c_A \approx$ 1100 km. This thickness estimation is of the same order as the optical depths, -5.3< *log* ($\tau_{500nm}$) <-2.0 (or 250-1100 km above the photosphere), over which Alfvén shocks are detected, highlighting the self-consistency of Alfvén shock formations in the lower solar atmosphere.

As a verification step, it is useful to track the plasma-β = 1 isocontours across the atmospheric heights (i.e., through the atmospheric thickness, L) where mode conversion and Alfvén shocks are detected. Computation of the plasma-β = 1 isocontours at different optical depths corresponding to *log* ($\tau_{500nm}$) = -2.0 (~250 km; where shocks are first detectable), *log* ($\tau_{500nm}$) = -3.0 (~450 km; more pronounced shock features) and *log* ($\tau_{500nm}$) = -5.0 (~1000 km; greatest shock temperature enhancements), reveal a gradual expansion of the mode conversion region with atmospheric height (Fig. S9). This is consistent with previous investigations[S10,S39] (see, also, Fig. S9) that identified an increase in the number of plasma-β < 1 locations with atmospheric height. Importantly, the contiguous coupling of plasma-β = 1 locations, as a function of atmospheric height, ensures that the sunspot atmosphere under study exhibits ideal conditions for the mode conversion of magneto-acoustic waves into Alfvén modes. It must be noted that only the base plasma-β = 1 isocontour (calculated for *log* ($\tau_{500nm}$) = -2.0 or ~250 km) is displayed in Fig. 1 in the main text. This is due to the fact that the base plasma-β = 1 isocontour is of the most significant importance, since the position of this inside the umbral atmosphere re-affirms the commencement of mode conversion between upwardly propagating umbral magneto-acoustic waves and their Alfvén mode counterparts. As a result, we did not wish to detract from the importance of this crucial location within the main text, while Fig. S9 now provides additional visual information for those wishing to further understand the evolution of the plasma-β = 1 isocontours with atmospheric height.

## 11. Future Studies of Alfvén Shocks using A-Priori Knowledge

The sunspot described here presents a unique magnetic field configuration, whereby the plasma-β = 1 region is co-located with the dark umbral perimeter. Here, magnetoacoustic-to-Alfvén mode conversion, and subsequent energy deposition through the creation of Alfvén shocks, is clearly visible when superimposed against the suppressed background emission of the sunspot umbra. However, not all sunspots display similar positioning of the plasma-β = 1 region, with stronger underlying sunspots pushing the equipartition layer towards its outer extremities as a result of the increased magnetic pressure[S48]. Does this reduce the efficiency of the mode conversion and subsequent Alfvén shock processes? Or does the enhanced emission towards the outer periphery of the sunspot penumbra mask the signatures of Alfvén shocks due to the higher background brightness? Furthermore, does the complexity of the umbral shape, enhanced through the presence of light bridges, mixed polarities, non-potential magnetic fields, etc., affect the rate at which Alfvén shocks can be generated? Finally, does the evolutionary timeframe of the sunspot contribute to the effectiveness of mode conversion when forming Alfvén shocks? For example, does decaying flux, synonymous with the final few days of a sunspot's lifetime, promote more structured magnetic field topologies that increase the spatial extent over which Alfvén shocks can be formed? Or do sunspots created during periods of heightened solar activity, which may naturally have elevated background flux concentrations[S96], help suppress the generation of Alfvén shocks within the boundaries of the active region due to the enhanced magnetic pressure?

To address these questions, future studies may employ full spectropolarimetric line profiles to examine the small-scale magnetic fluctuations also expected to manifest in Alfvén shocks[S60]. Inversion routines, applied to both photospheric and chromospheric spectra, will be able to better link the magnetic fields supporting mode conversion, thereby improving our understanding of which magnetic field topologies are required for efficient Alfvén shock dissipation. Furthermore, the improved spectral resolution provided by slit-based spectrographs will allow the morphology of the shocking plasma to be more thoroughly analysed through examination of spectral line widths (and associated broadening), higher Doppler precision and complete flux emission estimates. A statistical study, incorporating a vast number of umbrae across a complete solar cycle, will allow long-term studies of the role Alfvén shocks play in supplying heat to the active region plasma to be conclusively determined. Importantly, the results presented here are a pivotal turning point in our understanding of shock heating from MHD waves in the solar atmosphere. However, this knowledge is directly applicable to other fields of research, including laboratory plasma physics[S97] and high-energy astrophysics[S98].

**Supplementary References**

1. Cavallini, F. IBIS: A new post-focus instrument for solar imaging spectroscopy. *Sol. Phys.* **236**, 415-439 (2006).
2. Rimmele, T. R. Recent advances in solar adaptive optics. *Proc. SPIE* **5490**, 34-36 (2004).
3. Schou, J. et al. Design and ground calibration of the helioseismic and magnetic imager (HMI) instrument on the solar dynamics observatory (SDO). *Sol. Phys.* **275**, 229 (2012).
4. Pesnell, W. D., Thompson, B. J. & Chamberlin, P. C. The solar dynamics observatory (SDO). *Sol. Phys.* **275**, 3-15 (2012).
5. Borrero, J. M. et al. VFISV: Very fast inversion of the stokes vector for the helioseismic and magnetic imager. *Sol. Phys.* **273**, 267-293 (2011).
6. Gary, G. A. & Hagyard, M. J. Transformation of vector magnetograms and the problems associated with the effects of perspective and the azimuthal ambiguity. *Sol. Phys.* **126**, 21 (1990).
7. Wiegelmann, T. Nonlinear force-free modelling of the solar coronal magnetic field. *Journal of Geophysical Research (Space Physics)* **113**, A03S02 (2008).
8. Metcalf, T. R. et al. An overview of existing algorithms for resolving the 180° ambiguity in vector magnetic fields: Quantitative tests with synthetic data. *Sol. Phys.* **237**, 267-296 (2006).
9. Leka, K. D. et al. Resolving the 180° ambiguity in solar vector magnetic field data: Evaluating the effects of noise, spatial resolution, and method assumptions. *Sol. Phys.* **260**, 83 (2009).
10. Metcalf, T. R., Jiao, L., McClymont, A. N., Canfield, R. C. & Uitenbroek, H. Is the solar chromosphere magnetic field force-free? *Astrophys. J.* **439**, 474-481 (1995).
11. Aschwanden, M. J., Reardon, K. & Jess, D. B. Tracing the chromospheric and coronal magnetic field with AIA, IRIS, IBIS, and ROSA data. *Astrophys. J.* **826**, 61 (2016).
12. Socas-Navarro, H. The three-dimensional structure of a sunspot magnetic field. *Astrophys. J.* **631**, L167- L170 (2005).
13. Aschwanden, M. J. The vertical-current approximation nonlinear force-free field code - Description, performance tests, and measurements of magnetic energies dissipated in solar flares *Astrophys. J. Supp.* **224**, 25 (2016).
14. Aschwanden, M., De Pontieu, B. & Katrukha, E. Optimization of curvilinear tracing applied to solar physics and biophysics. *Entropy* **15**, 3007-3030 (2013).
15. Jess, D. B. et al. The source of 3 minute magnetoacoustic oscillations in coronal fans. *Astrophys. J.* **757**, 160 (2012).
16. Aschwanden, M. J. Nonlinear force-free magnetic field fitting to coronal loops with and without spectroscopy. *Astrophys. J.* **763**, 115 (2013).
17. Maltby, P. et al. A new sunspot umbral model and its variation with the solar cycle. *Astrophys. J.* **306**, 284-303 (1986).
18. Beckers, J. M. & Tallant, P. E. Chromospheric inhomogeneities in sunspot umbrae. *Sol. Phys.* **7**, 351-365 (1969).
19. Bharti, L., Solanki, S. K. & Hirzberger, J. Lambda-shaped jets from a penumbral intrusion into a sunspot umbra: a possibility for magnetic reconnection. *Astron. Astrophys.* **597**, A127 (2017).
20. Yurchyshyn, V., Abramenko, V., Kosovichev, A. & Goode, P. High resolution observations of chromospheric jets in sunspot umbra. *Astrophys. J.* **787**, 58 (2014).

21. Vissers, G. J. M., Rouppe van der Voort, L. H. M. & Carlsson, M. Evidence for a transition region response to penumbral microjets in sunspots. *Astrophys. J.* **811**, L33 (2015).
22. Katsukawa, Y. et al. Small-scale jetlike features in penumbral chromospheres. *Science* **318**, 1594 (2007).
23. Kirichenko, A. S. & Bogachev, S. A. Plasma heating in solar microflares: Statistics and analysis. *Astro- phys. J.* **840**, 45 (2017).
24. Hong, J. et al. Bidirectional outflows as evidence of magnetic reconnection leading to a solar microflare. *Astrophys. J.* **820**, L17 (2016).
25. Vecchio, A., Cauzzi, G. & Reardon, K. P. The solar chromosphere at high resolution with IBIS. II. Acoustic shocks in the quiet internetwork and the role of magnetic fields. *Astron. Astrophys.* **494**, 269- 286 (2009).
26. Georgoulis, M. K., Tziotziou, K. & Raouafi, N.-E. Magnetic energy and helicity budgets in the active- region solar corona. II. Nonlinear force-free approximation. *Astrophys. J.* **759**, 1 (2012).
27. Georgoulis, M. K. & LaBonte, B. J. Magnetic energy and helicity budgets in the active region solar corona. I. Linear force-free approximation. *Astrophys. J.* **671**, 1034-1050 (2007).
28. Aschwanden, M. J., Xu, Y. & Jing, J. Global energetics of solar flares. I. Magnetic energies. *Astrophys. J.* **797**, 50 (2014).
29. Cauzzi, G. et al. The solar chromosphere at high resolution with IBIS. I. New insights from the Ca II 854.2 nm line. *Astron. Astrophys.* **480**, 515-526 (2008)
30. Wiegelmann, T., Thalmann, J. K. & Solanki, S. K. The magnetic field in the solar atmosphere. *Astronomy and Astrophysics Reviews* **22**, 78 (2014).
31. Moon, Y.-J., Choe, G. S., Yun, H. S., Park, Y. D. & Mickey, D. L. Force-freeness of solar magnetic fields in the photosphere. *Astrophys. J.* **568**, 422-431 (2002).
32. Georgoulis, M. K. & LaBonte, B. J. Vertical lorentz force and cross-field currents in the photospheric magnetic fields of solar active regions. *Astrophys. J.* **615**, 1029-1041 (2004).
33. Régnier, S. & Priest, E. R. Nonlinear force-free models for the solar corona. I. Two active regions with very different structure. *Astron. Astrophys.* **468**, 701-709 (2007).
34. Mathew, S. K. et al. Thermal-magnetic relation in a sunspot and a map of its Wilson depression. *Astron. Astrophys.* **422**, 693-701 (2004).
35. Metcalf, T. R. et al. Nonlinear force-free modeling of coronal magnetic fields. II. Modeling a filament arcade and simulated chromospheric and photospheric vector fields. *Sol. Phys.* **247**, 269-299 (2008).
36. De Rosa, M. L. et al. A critical assessment of nonlinear force-free field modeling of the solar corona for active region 10953. *Astrophys. J.* **696**, 1780-1791 (2009).
37. Liu, S. et al. A statistical study on force-freeness of solar magnetic fields in the photosphere. *Publications of the Astron. Soc. of Australia* **30**, e005 (2013). 25
38. Sandman, A. W., Aschwanden, M. J., Derosa, M. L., Wülser, J. P. & Alexander, D. Comparison of STEREO/EUVI loops with potential magnetic field models. *Sol. Phys.* **259**, 1-11 (2009).
39. Gary, G. A. Plasma beta above a solar active region: Rethinking the paradigm. *Sol. Phys.* **203**, 71 (2001).
40. Wiegelmann, T. et al. Magneto-static modeling of the mixed plasma beta solar atmosphere based on Sunrise/IMaX data. *Astrophys. J.* **815**, 10 (2015).
41. Wiegelmann, T. et al. Magneto-static modeling from Sunrise/IMaX: Application to an active region observed with Sunrise II. *Astrophys. J. Supp.* **229**, 18 (2017).
42. Wiegelmann, T. & Inhester, B. How to deal with measurement errors and lacking data in nonlinear force-free coronal magnetic field modelling? *Astron. Astrophys.* **516**, A107 (2010).
43. Wiegelmann, T. et al. How should one optimize nonlinear force-free coronal magnetic field extrapolations from SDO/HMI vector magnetograms? *Sol. Phys.* **281**, 37-51 (2012).
44. Centeno, R., Trujillo Bueno, J. & Asensio Ramos, A. On the magnetic field of off-limb spicules. *Astrophys. J.* **708**, 1579-1584 (2010).

45. Schad, T. A., Penn, M. J. & Lin, H. He I vector magnetometry of field-aligned superpenumbral fibrils. *Astrophys. J.* **768**, 111 (2013).
46. Wiegelmann, T., Petrie, G. J. D. & Riley, P. Coronal magnetic field models. *Space Sci. Rev.* **210**, 249-274 (2017).
47. Kobanov, N. I., Kolobov, D. Y. & Makarchik, D. V. Umbral three-minute oscillations and running penumbral waves. *Sol. Phys.* **238**, 231-244 (2006).
48. Jess, D. B., Reznikova, V. E., Van Doorsselaere, T., Keys, P. H. & Mackay, D. H. The influence of the magnetic field on running penumbral waves in the solar chromosphere. *Astrophys. J.* **779**, 168 (2013).
49. Grant, S. D. T. et al. Wave damping observed in upwardly propagating sausage-mode-oscillations contained within a magnetic pore. *Astrophys. J.* **806**, 132 (2015).
50. Beckers, J. M. & Schröter, E. H. The intensity, velocity and magnetic structure of a sunspot region. II: Some properties of umbral dots. *Sol. Phys.* **4**, 303-314 (1968).
51. Rouppe van der Voort, L. H. M., Rutten, R. J., Sütterlin, P., Sloover, P. J. & Krijger, J. M. La Palma observations of umbral flashes. *Astron. Astrophys.* **403**, 277-285 (2003).
52. Madsen, C. A., Tian, H. & DeLuca, E. E. Observations of umbral flashes and running sunspot waves with the interface region imaging spectrograph. *Astrophys. J.* **800**, 129 (2015).
53. Norton, A. A., Jones, E. H. & Liu, Y. How do the magnetic field strengths and intensities of sunspots vary over the solar cycle. *Journal of Physics Conference Series* **440**, 012038 (2013).
54. Khomenko, E. & Collados, M. Oscillations and waves in sunspots. *Living Reviews in Solar Physics* **12**, 6 (2015).
55. Ferraro, V. C. A. Hydromagnetic waves in a rare ionized gas and galactic magnetic fields. *Proceedings of the Royal Society of London Series A* **233**, 310-318 (1955).
56. Medvedev, M. V. Collisionless dissipative nonlinear Alfvén waves: Nonlinear steepening, compressible turbulence, and particle trapping. *Physics of Plasmas* **6**, 2191-2197 (1999).
57. Mikhailovskii, A. B., Petviashvili, V. I. & Fridman, A. M. The Alfvén soliton. *ZhETF Pisma Redaktsiiu* **24**, 53-56 (1976).
58. Khomenko, E., Collados, M., Díaz, A. & Vitas, N. Fluid description of multi-component solar partially ionized plasma. *Physics of Plasmas* **21**, 092901 (2014).
59. Cohen, R. H. & Kulsrud, R. M. Nonlinear evolution of parallel-propagating hydromagnetic waves. *Physics of Fluids* **17**, 2215-2225 (1974).
60. Montgomery, D. Development of hydromagnetic shocks from large-amplitude Alfvén waves. *Phys. Rev. Lett.* **2**, 36-37 (1959).
61. Hollweg, J. V., Jackson, S. & Galloway, D. Alfvén waves in the solar atmosphere. III - Nonlinear waves on open flux tubes. *Sol. Phys.* **75**, 35-61 (1982).
62. Braginskii, S. I. Transport processes in a plasma. *Reviews of Plasma Physics* **1**, 205 (1965).
63. Zaqarashvili, T. V., Khodachenko, M. L. & Soler, R. Torsional Alfvén waves in a partially ionized solar plasma: Effects of neutral helium and stratification. *Astron. Astrophys.* **549**, A113 (2013).
64. Hollweg, J. V. Density fluctuations driven by Alfvén waves. *J. Geophys. Res.* **76**, 5155 (1971).
65. Zaqarashvili, T. V., Oliver, R. & Ballester, J. L. Spectral line with decrease in the solar corona: Resonant energy conversion from Alfvén to acoustic waves. *Astron. Astrophys.* **456**, L13-L16 (2006).
66. Socas-Navarro, H., Trujillo Bueno, J. & Ruiz Cobo, B. Anomalous circular polarization profiles in sunspot chromospheres. *Astrophys. J.* **544**, 1141-1154 (2000).
67. Kerr, G. S., Simões, P. J. A., Qiu, J. & Fletcher, L. IRIS observations of the Mg II h and k lines during a solar flare. *Astron. Astrophys.* **582**, A50 (2015).
68. Beckers, J. M. & Schultz, R. B. Oscillatory motions in sunspots. *Sol. Phys.* **27**, 61-70 (1972).
69. Giovanelli, R. G., Harvey, J. W. & Livingston, W. C. Motions in solar magnetic tubes. III - Outward wave propagation in sunspot umbras. *Sol. Phys.* **58**, 347-361 (1978).
70. Tian, H. et al. High-resolution observations of the shock wave behaviour for sunspot oscillations with the interface region imagins spectrograph. *Astrophys. J.* **786**, 137 (2014).

71. Rezaei, R. & Beck, C. Multiwavelength spectropolarimetric observations of an Ellerman bomb. *Astron. Astrophys.* **582**, A104 (2015).
72. Beck, C., Rezaei, R. & Puschmann, K. G. The energy of waves in the photosphere and lower chromo- sphere. III. Inversion setup for Ca II H spectra in local thermal equilibrium. *Astron. Astrophys.* **549**, A24 (2013).
73. Beck, C., Rezaei, R. & Puschmann, K. G. The energy of waves in the photosphere and lower chromo- sphere. IV. Inversion results of Ca II H spectra. *Astron. Astrophys.* **553**, A73 (2013).
74. Beck, C., Choudhary, D. P. & Rezaei, R. A three-dimensional view of the thermal structure in a super- penumbral canopy. *Astrophys. J.* **788**, 183 (2014).
75. Beck, C., Choudhary, D. P., Rezaei, R. & Louis, R. E. Fast inversion of solar Ca II spectra. *Astrophys. J.* **798**, 100 (2015).
76. Pearson, K. On the criterion that a given system of deviations from the probable in the case of a correlated system of variables is such that it can be reasonably supposed to have arisen from random sampling. *Philosophical Magazine Series 5* **50**, 157-175 (1900).
77. Pfeiffer, P. E. & Schum, D. A. *Introduction to Applied Probability* (Academic Press, New York, 1973).
78. Bel, N. & Leroy, B. Analytical study of magnetoacoustic gravity waves. *Astron. Astrophys.* **55**, 239 (1977).
79. de la Cruz Rodríguez, J., Rouppe van der Voort, L., Socas-Navarro, H. & van Noort, M. Physical properties of a sunspot chromosphere with umbral flashes. *Astron. Astrophys.* **556**, A115 (2013).
80. Socas-Navarro, H., Trujillo Bueno, J. & Ruiz Cobo, B. Anomalous polarization profiles in sunspots: Possible origin of umbral flashes. *Science* **288**, 1396-1398 (2000).
81. Socas-Navarro, H., Trujillo Bueno, J. & Ruiz Cobo, B. A time-dependent semiempirical model of the chromospheric umbral oscillation. *Astrophys. J.* **550**, 1102-1112 (2001).
82. Centeno, R., Socas-Navarro, H., Collados, M. & Trujillo Bueno, J. Evidence for fine structure in the chromospheric umbral oscillation. *Astrophys. J.* **635**, 670-673 (2005).
83. Bard, S. & Carlsson, M. Radiative hydrodynamic simulations of acoustic waves in sunspots. *Astrophys. J.* **722**, 888 (2010).
84. Felipe, T., Socas-Navarro, H. & Khomenko, E. Synthetic observations of wave propagation in a sunspot umbra. *Astrophys. J.* **795**, 9 (2014).
85. Socas-Navarro, H., de la Cruz Rodríguez, J., Asensio Ramos, A., Trujillo Bueno, J. & Ruiz Cobo, B. An open-source, massively parallel code for non-LTE synthesis and inversion of spectral lines and Zeeman induced Stokes profiles. *Astron. Astrophys.* **577**, A7 (2015).
86. De la Cruz Rodríguez, J., Socas-Navarro, H., Carlsson, M. & Leenaarts, J. Non-local thermodynamic equilibrium inversions from a 3D magnetohydrodynamic chromospheric model. *Astron. Astrophys.* **543**, A34 (2012).
87. Vranjes, J., Poedts, S., Pandey, B. P. & de Pontieu, B. Energy flux of Alfvén waves in weakly ionized plasma. *Astron. Astrophys.* **478**, 553-558 (2008).
88. Zaqarashvili, T. V. & Roberts, B. Two-wave interaction in ideal magnetohydrodynamics. *Astron. Astrophys.* **452**, 1053-1058 (2006).
89. Cally, P. S. What to look for in the seismology of solar active regions. *Astronomische Nachrichten* **328**, 286 (2007).
90. Krishna Prasad, S., Jess, D. B. & Khomenko, E. On the source of propagating slow magnetoacoustic waves in sunspots. *Astrophys. J.* **812**, L15 (2015).
91. Jess, D. B. et al. Solar coronal magnetic fields derived using seismology techniques applied to omnipresent sunspot waves. *Nature Phys.* **12**, 179-185 (2016).
92. Cally, P. S. & Goossens, M. Three-dimensional MHD wave propagation and conversion to Alfvén waves near the solar surface. I. Direct numerical solution. *Sol. Phys.* **251**, 251-265 (2008).
93. Cho, I.-H. et al. Determination of the Alfvén speed and plasma-beta using the seismology of sunspot umbra. *Astrophys. J.* **837**, L11 (2017).

94. Savitzky, A. & Golay, M. J. E. Smoothing and differentiation of data by simplified least squares procedures. *Analytical Chemistry* **36**, 1627-1639 (1964).
95. Steinier, J., Termonia, Y. & Deltour, J. Smoothing and differentiation of data by simplified least squares procedures. *Analytical Chemistry* **44**, 1906-1909 (1972).
96. Watson, F. T., Fletcher, L. & Marshall, S. Evolution of sunspot properties during solar cycle 23. *Astron. Astrophys.* **533**, A14 (2011).
97. Cairns, R. A. & Fuchs, V. Mode conversion and dissipation of the fast Alfvén wave in ion cyclotron heating. *Physics of Fluids B* **1**, 350-357 (1989).
98. Punsly, B., Balsara, D., Kim, J. & Garain, S. Riemann solvers and Alfvén waves in black hole magnetospheres. *Computational Astrophysics and Cosmology* **3**, 5 (2016).